\documentclass[10pt,journal]{article}

\usepackage[english]{babel}
\usepackage[T1]{fontenc}
\usepackage{amssymb}
\usepackage{amsfonts}
\usepackage{amsmath}
\usepackage{amsthm}
\usepackage{mathpazo}
\usepackage{bm}
\usepackage[colorinlistoftodos]{todonotes}
\usepackage{soul}
\usepackage{enumitem}
\usepackage{float}
\usepackage{rotating}
\usepackage{multirow}
\usepackage[utf8]{inputenc} 
\usepackage{algorithm}
\usepackage{algpseudocode}
\usepackage[colorlinks=true, allcolors=blue]{hyperref}
\usepackage{lipsum,graphicx}
\usepackage{hhline}
\usepackage{setspace}
\usepackage[top=2.54cm,bottom=2.54cm, left=2.54cm, right=2.54cm]{geometry}
\usepackage{epstopdf}
\usepackage{xcolor}

\title{A Monte Carlo comparison of categorical tests of independence}
\author{Abdulaziz Alenazi \\
Department of Mathematics, Northern Border University, Arar, Saudi Arabia, \\ \href{mailto:a.alenazi@nbu.edu.sa}{a.alenazi@nbu.edu.sa} \\
}

\begin{document}

\maketitle

\begin{center}
{\bf Abstract}
\end{center}
The $X^2$ and $G^2$ tests are the most frequently applied tests  for testing the independence of two categorical variables. However, no one, to the best of our knowledge has compared them, extensively, and ultimately answer the question of which to use and when. Further, their applicability in cases with zero frequencies has been debated and (non parametric) permutation tests are suggested. In this work we perform extensive Monte Carlo simulation studies attempting to answer both aforementioned points. As expected, in large sample sized cases ($>1,000$) the $X^2$ and $G^2$ are indistinguishable. In the small sample sized cases ($\leq 1,000$) though, we provide strong evidence supporting the use of the $X^2$ test regardless of zero frequencies for the case of unconditional independence. Also, we suggest the use of the permutation based $G^2$ test for testing conditional independence, at the cost of being computationally more expensive. The $G^2$ test exhibited inferior performance and its use should be limited. 
\vspace{18pt}


\section{Introduction}		
Categorical data are frequently encountered in many disciplines outside statistics, mainly in the ones that fall under the social sciences umbrella, but also in medicine, biology, sociology, psychology, computer science, political sciences, demography, bioinformatics, dentistry, geology, etc. These data can either be ordinal or nominal variables. Specifically with nominal categorical variables, for which independence is to be tested, the $X^2$ test and the $G^2$ test are two commonly applicable tests, both of which are able to capture non-linear relationships between the variables. Even though the list of independence tests is wider \cite{williams1976}, \cite{rao1981}, \cite{rao1984}, the aforementioned tests have gained popularity mainly due to their simplicity.  

Both tests are considered to be non-parametric as they make no assumption of the underlying distribution of the data. They do however entail some assumptions as stated by \cite{mchugh2013}. Prior to their application one must first produce a contingency table and as \cite{mchugh2013} mentions, the cells should contain frequencies of pairs of values and not percentages. The values of the variables must be mutually exclusive. For instance, in a psychology experiment the respondent can choose only one answer. In the same experiment, all respondents must have answered the questions only once, that is, repeated measurements should not be analyzed using this test. \cite{mchugh2013} states that the categorical variables can also be ordinal, e.g. nominal-ordinal or ordinal-ordinal. A necessary condition applied to the expected values should be satisfied. The number of cells whose expected values is less than 5 should be no more than 20\%-25\% and no cell must contain 0 frequencies. According to \cite{mchugh2013}, no cell should contain expected values that are less than 3.  

In a $2\times2$ contingency table for example with low cell values and or zeros, Fisher's exact test (\cite{fisher1922}) is ordinarily, even though researchers have argued against its use in such cases (\cite{liddell1976}, \cite{berkson1978} ) because its actual rejection rate is below the set nominal significance level. Nonetheless, researchers have proposed generalizations to $r \times c$ contingency tables (\cite{wells1980}, \cite{mehta1983}). Another alternative is to compute the associated p-value using Monte Carlo simulation or random permutations (\cite{tsamardinos2010}) which is applicable to larger than $2\times2$ contingency tables as well. 

The convenience and ease of application of those categorical tests of independence has made them widely applicable in the social sciences ( \cite{wilcox1996}), but also in the computer science and bioinformatics for network construction ( \cite{tsamardinos2006}). Despite them being so broadly used, no one, to the best of our knowledge, has performed simulation studies in order to give guidance as to which test to use and when. One exception could be \cite{tsamardinos2010} who showed that the permutations based $G^2$ improves the learning quality of Bayesian networks. Their considered though only the $G^2$ with and without permutations.  

The scope of this paper is to provide evidence that the percentage of cells that contain expected values less than 5, is incorrect and should be neglected. In addition, we provide evidence as to when to compute the asymptotic p-value of the $X^2$ or of the $G^2$ and when to rely on the permutation based p-value. The evidence is based on simulation studies we conducted covering multiple scenarios of simple independence and of conditional independence. The simulation studies showed that the $G^2$ performs bad in terms attainment of the type I error, whereas the $X^2$ test performed better but is not always applicable. The permutations based $G^2$ though, produced satisfactorily results, but only in certain cases. 

The next section presents the two aforementioned tests covering both unconditional and conditional independence. Section 3 presents the simulation studies, and section 4 concludes the paper.	

\section{Tests of independence for categorical data}
Suppose we have $n$ observations from two categorical variables $X$ and $Y$ taking discrete values and denote by $|X|$ and $|Y|$ their cardinalities. The ordinary null and alternative hypotheses are
\begin{eqnarray*}
& & H_0: X \ \text{and} \ Y \ \text{are independent} \\
& & H_1: X \ \text{and} \ Y \ \text{are \textbf{not} independent}
\end{eqnarray*}
The most famous independence tests for testing the above null hypothesis are the $X^2$ and the $G^2$ tests. 

\subsection{The $X^2$ test of (unconditional) independence}
The $X^2$ test was proposed by \cite{pearson1900} and its test statistic is given by
\begin{eqnarray} \label{chi2}
X^2(X, Y)=\sum_{i, j}\frac{\left(O_{ij} - E_{ij}\right)^2}{E_{ij}^2}
\end{eqnarray}
The $O_{ij}$ are the observed frequencies of the $i-th$ and $j-th$ values of $X$ and $Y$ respectively. The $E_{ij}$ are their corresponding expected frequencies computed as $E_{ij}=\frac{O_{i+}O_{+j}}{O_{++}}$, where $O_{i+} = \sum_{j=1}^nO_{ij}$, $O_{+j}=\sum_{i=1}^nO_{ij}$ and $O_{++}=n$. Under the independence assumption, $X^2 \sim \chi^2_{(|X| - 1) (|Y| - 1)}$.

\subsection{The $G^2$ test of  (unconditional) independence}
The $G^2$ test or $G$-test is simply a log-likelihood ratio test whose statistic is given by \cite{agresti2002}
\begin{eqnarray} \label{g2}
G^2(X, Y)=2\sum_{i, j}O_{ij}\log{\frac{O_{ij}}{E_{ij}}}
\end{eqnarray}
Similarly to the $X^2$ test, under $H_0$,  $G^2 \sim \chi^2_{(|X| - 1) (|Y| - 1)}$.

\subsection{The $X^2$ and $G^2$-tests of conditional independence}
When we wish to test whether the two variables are independent conditional on one or more variables {\bf Z} null and alternative hypotheses are
\begin{eqnarray*}
& & H_0: X \ \text{and} \ Y \ \text{are independent conditional on $\bf Z$} \\
& & H_1: X \ \text{and} \ Y \ \text{are \textbf{not} independent conditional on $\bf Z$}
\end{eqnarray*}
and the $X^2$ and $G^2$ test statistics are given by \cite{agresti2002}
\begin{eqnarray} \label{condtests}
\begin{array}{cccc}
X^2(X, Y|{\bf Z}) &  =  & \sum_{i, j}\frac{\left(O_{ij|k} - E_{ij|k}\right)^2}{E_{ij|k}^2} &  \text{and} \\
G^2(X, Y|{\bf Z}) & = & 2\sum_k\sum_{i, j}O_{ij|k}\log{\frac{O_{ij|k}}{E_{ij|k}}}  & \text{respectively,}
\end{array}
\end{eqnarray}
where $k$ denotes the $k$-th value of ${\bf Z}$ and $|{\bf Z}|$ indicates the cardinality of ${\bf Z}$, the total number of values of ${\bf Z}$. The $O_{ij}$ are the observed frequencies of the $i-th$ and $j-th$ values of $X$ and $Y$ respectively for the $k$-th value of $\bf Z$. The $E_{ij}$ are their corresponding expected frequencies computed as $E_{ij}=\frac{O_{i+|k}O_{+j|k}}{O_{++|k}}$, where $O_{i+|k} = \sum_{j=1}^nO_{ij|k}$, $O_{+j|k}=\sum_{i=1}^nO_{ij|k}$ and $O_{++|k}=n_k$. Under the independence assumption, $X^2 \sim \chi^2_{(|X| - 1) (|Y| - 1)}$. Under $H_0$, both $X^2$ and $G^2$ follow a $\chi^2_{(|X| - 1) (|Y| - 1)|{\bf Z}|}$. It becomes clear that (\ref{chi2}) and (\ref{g2}) are special cases of (\ref{condtests}) when ${\bf Z} = \emptyset$. 

\subsection{Permutation based p-values}
The aforementioned test statistics produce asymptotic p-values. Computer intensive methods include Monte Carlo simulations and permutations. In this paper we will rely on permutations to obtain the p-value. With continuous variables for example, the idea is to distort the pairs multiple times and each time calculate the relevant test statistic (based on Pearson or Spearman). With categorical variables though, extra caution must be taken. Similarly to Fisher's exact test (\cite{agresti2002}) the permutations must occur in such a way as to keep the row and column totals fixed. The p-value is then computed as the proportion of times the values of the permuted test statistics exceed the value of the test statistic in the original data.

\subsection{Relationship between the $X^2$ and the $G^2$ test}
The $X^2$ test statistic is an approximation of the $G^2$ using a second order Taylor expansion of the natural logarithm around $0$ (\cite{hoey2012}). Following \cite{hoey2012} let us write $O_{ij}=E_{ij}+\tau_{ij}$ and since $\sum_{i,j}O_{ij} =\sum_{i,j}E_{ij}$ this implies that $\sum_{i,j}\tau_{ij}=0$. The $G^2$ test (\ref{g2}) is then given by
\begin{eqnarray*}
G^2 &=& 2\sum_{i, j}O_{ij}\log{\frac{O_{ij}}{E_{ij}}} = 2\sum_{i, j}\left(E_{ij}+\tau_{ij}\right)\log{\frac{E_{ij}+\tau_{ij}}{E_{ij}}} = 2\sum_{i, j}\left(E_{ij}+\tau_{ij}\right)\log{\left(1 + \frac{\tau_{ij}}{E_{ij}}\right)} \\
& & \text{By expanding $\frac{\tau_{ij}}{E_{ij}}$ around $0$ we obtain} \\
G^2 &\approx& 2\sum_{i, j}\left(E_{ij}+\tau_{ij}\right)\left[\frac{\tau_{ij}}{E_{ij}} - \frac{1}{2}\left(\frac{\tau_{ij}}{E_{ij}}\right)^2+O\left(\tau_{ij}^3\right) \right] \\
G^2 &=& 2\sum_{ij}\tau_{ij} - \sum_{ij}\frac{\tau^2_{ij}}{E_{ij}} + O\left(\tau_{ij}^3\right) +
2\sum_{ij}\frac{\tau^2_{ij}}{E_{ij}} - \frac{1}{2}\sum_{ij}\frac{\tau^3_{ij}}{E_{ij}} + O\left(\tau_{ij}^4\right) \\
G^2 &=& \sum_{ij}\frac{\tau^2_{ij}}{E_{ij}} +O\left(\tau_{ij}^3\right) \approx \sum_{ij}\frac{\left(O_{ij}-E_{ij}\right)^2}{E_{ij}}
\end{eqnarray*}
\cite{hoey2012}, \cite{hoey2012} mentions that as the difference between $O_{ij}$ and $E_{ij}$ increases, the less accurate the above approximation becomes, and $X^2$ will tend to compute erroneous answers, especially in the small sample sized data. Further, since $X^2$ is an approximation to the $G^2$, the former is expected to be less accurate the latter. 

\section{Simulation studies}
We used R 3.6 (\cite{R2019}, \cite{R2019}) and the library \textit{Rfast} (\cite{rfast2019}) that contains fast implementations of the aforementioned testing procedures. R's built in function \textit{chisq.test} can return a Monte Carlo p-value, but not a permutation p-value, and it is very slow when thousands of tests must be conducted. \cite{tsagris2017} compared some implementations in R, showing that conditional independence testing is faster via Poisson log-linear models. Tests of independence also exist in the R package \textit{coin} (\cite{coin2008}) but are not as computationally efficient as the ones in the R package \textit{Rfast}. Computational efficiency was the reason why we did not consider a computer intensive version of the $X^2$ test and we only used the tests that are available in \textit{Rfast}.  

Table \ref{time} presents an example of the computational cost of each testing procedure. The $G^2$ test is slightly slower than the $X^2$ test because it involves computation of logarithms. The permutation $G^2$ test, that performs $999$ permutations and computes $999$ test statistics, is remarkably fast, as it is at most 2 times slower than R's built in $X^2$ test which performs only a single test. R's built in \textit{chisq.test} offers the option of a Monte Carlo p-value, but that would be extremely show for our example here. A reasonable estimate of the computational cost of this Monte Carlo p-value would be to multiply the last column of Table \ref{time} with 999, resulting in more than half an hour.  

\begin{table}[!ht]
\caption{\textbf{Time (in seconds)} required by the testing procedures to perform 4950 tests for different sample sizes and cardinalities. The relative computational cost, normalised with response to the $X^2$ test, appears inside the parentheses. \label{time}}
\begin{center}
\begin{tabular}{cc|cccc} \hline \hline
Sample size &  Cardinalities & $X^2$ test  &  $G^2$ test  & Permutation   & R's built in               \\ 
                   &                       &                     &                      & $G^2$ test    & $X^2$ test              \\     \hline \hline
n = 100 & $|X| = |Y| = 2$ & 0.003(1.00)  &  0.004(1.12)  &  0.867(256.96)  &  2.521(746.95)  \\
n = 200 & $|X| = |Y| = 3$ & 0.005(1.00)  &  0.006(1.23)  &  2.655(549.11)  &  2.437(504.02) \\
n = 400 & $|X| = |Y| = 4$ & 0.007(1.00)  &  0.010(1.48)  &  5.962(860.82)  &  3.031(437.59) \\
n = 800 & $|X| = |Y| = 5$ & 0.012(1.00)  &  0.019(1.50)  &  6.645(536.00)  &  3.426(276.37) \\     \hline \hline
n = 10,000 & $|X| = |Y| = 2$ & 0.141(1.00)  &  0.136(0.96)  &  3.369(23.87)  &  9.558(67.71)  \\
n = 10,000 & $|X| = |Y| = 3$ &  0.149(1.00)  &  0.135(0.91)  &  6.576(44.24)  &  10.015(67.38) \\
n = 10,000 & $|X| = |Y| = 4$ &  0.146(1.00)  &  0.144(0.99)  &  10.763(73.93)  &  10.355(71.12) \\
n = 10,000 & $|X| = |Y| = 5$ &  0.14(1.00)  &  0.138(0.99)  &  10.7(76.51)  &  10.165(72.68) \\
\hline \hline
\end{tabular}
\end{center}
\end{table}

\subsection{Difference between the $G^2$ and the $X^2$ test statistics}
\begin{figure}[ht]
\centering
\begin{tabular}{ccc}
\multicolumn{3}{c}{\underline{Sample sizes up to $1,000$, increasing by a step equal to $20$.}} \\
\includegraphics[scale = 0.32, trim = 50 0 0 0]{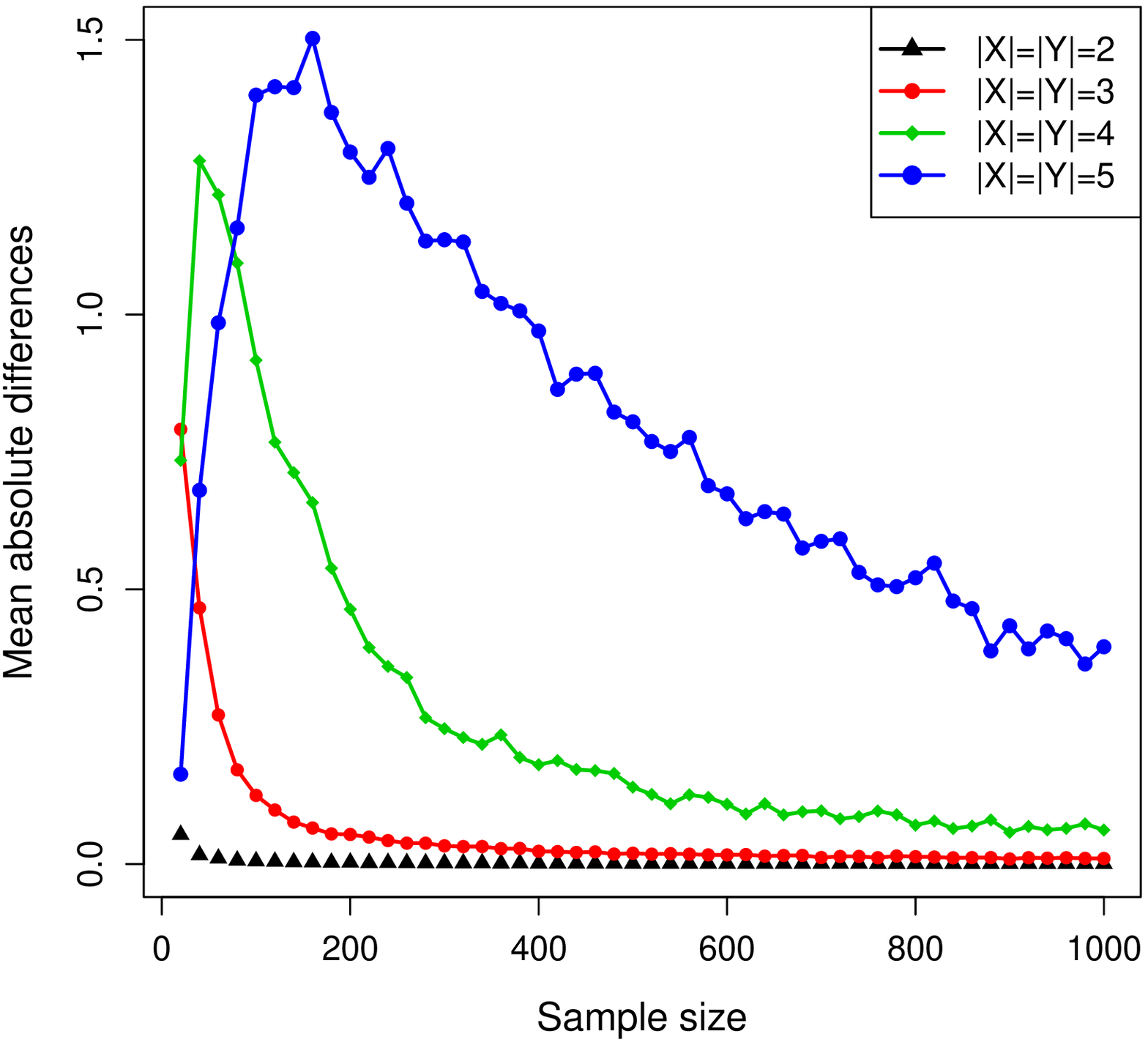} &
\includegraphics[scale = 0.32, trim = 35 0 0 0]{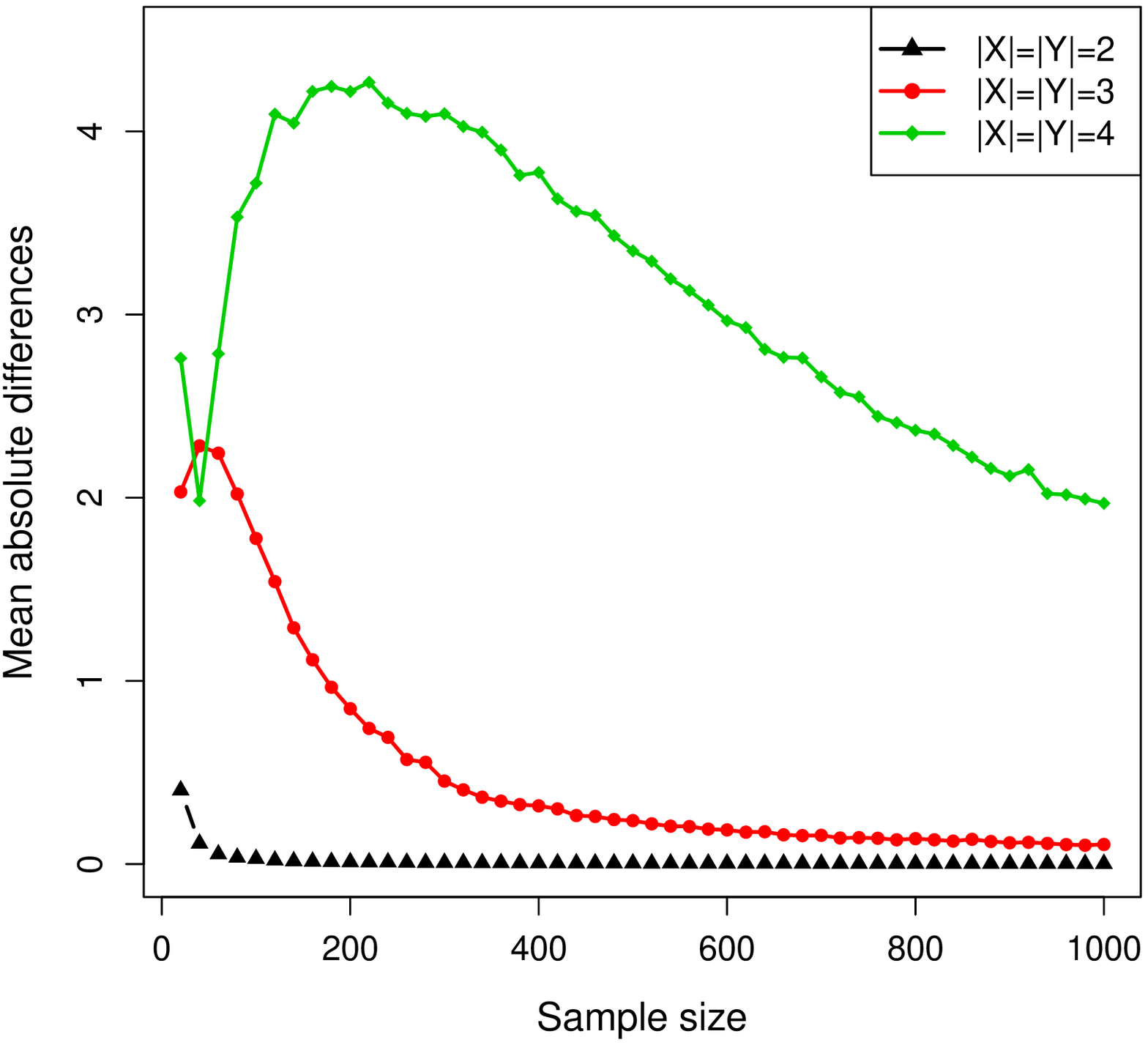} & 
\includegraphics[scale = 0.32, trim = 45 0 0 0]{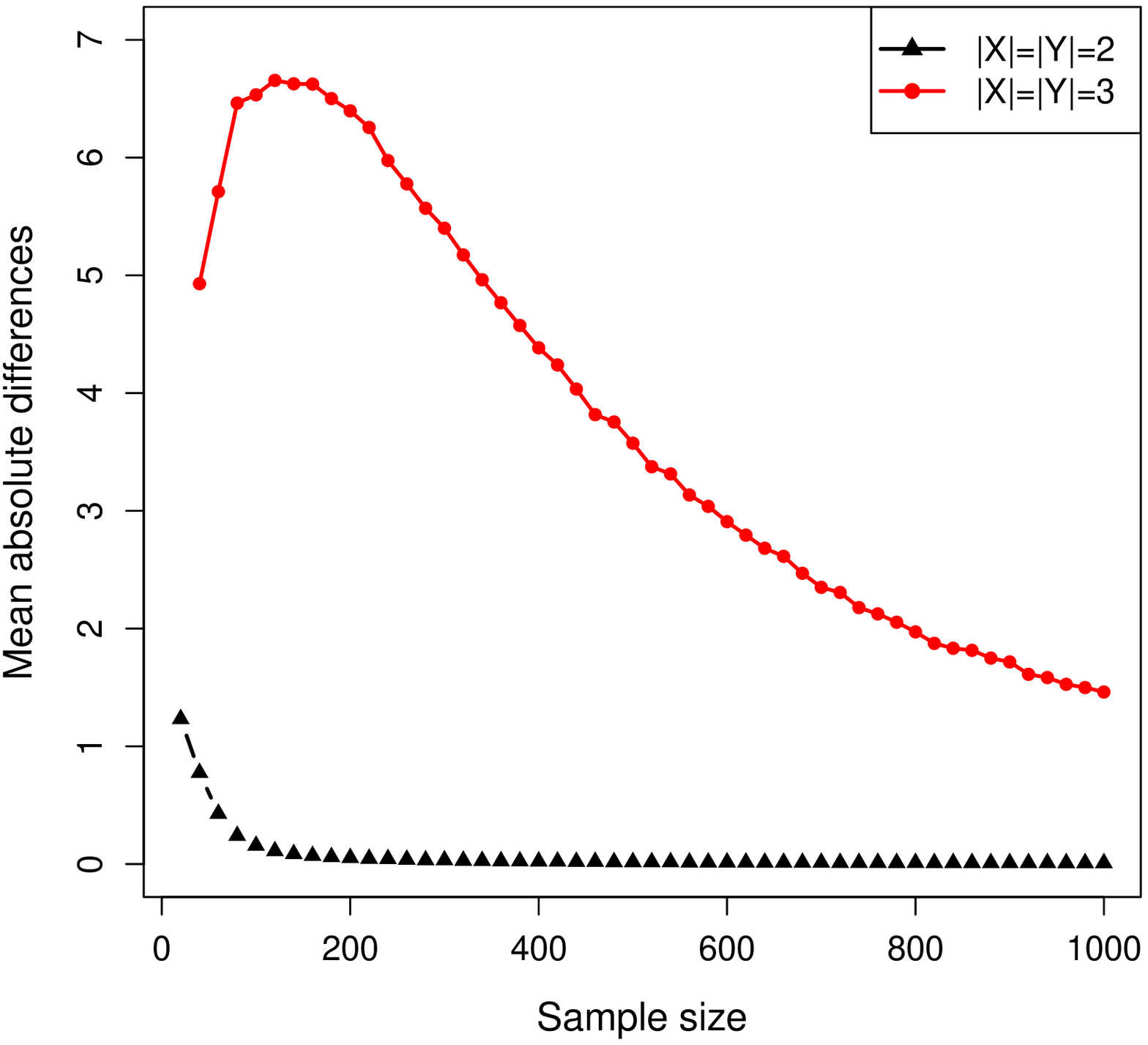} \\
(a)    &    (b)    &    (c)   \\
\multicolumn{3}{c}{\underline{Sample sizes, from $1,000$ up to $10,000$, increasing by $1,000$.}} \\
\includegraphics[scale = 0.32, trim = 50 0 0 0]{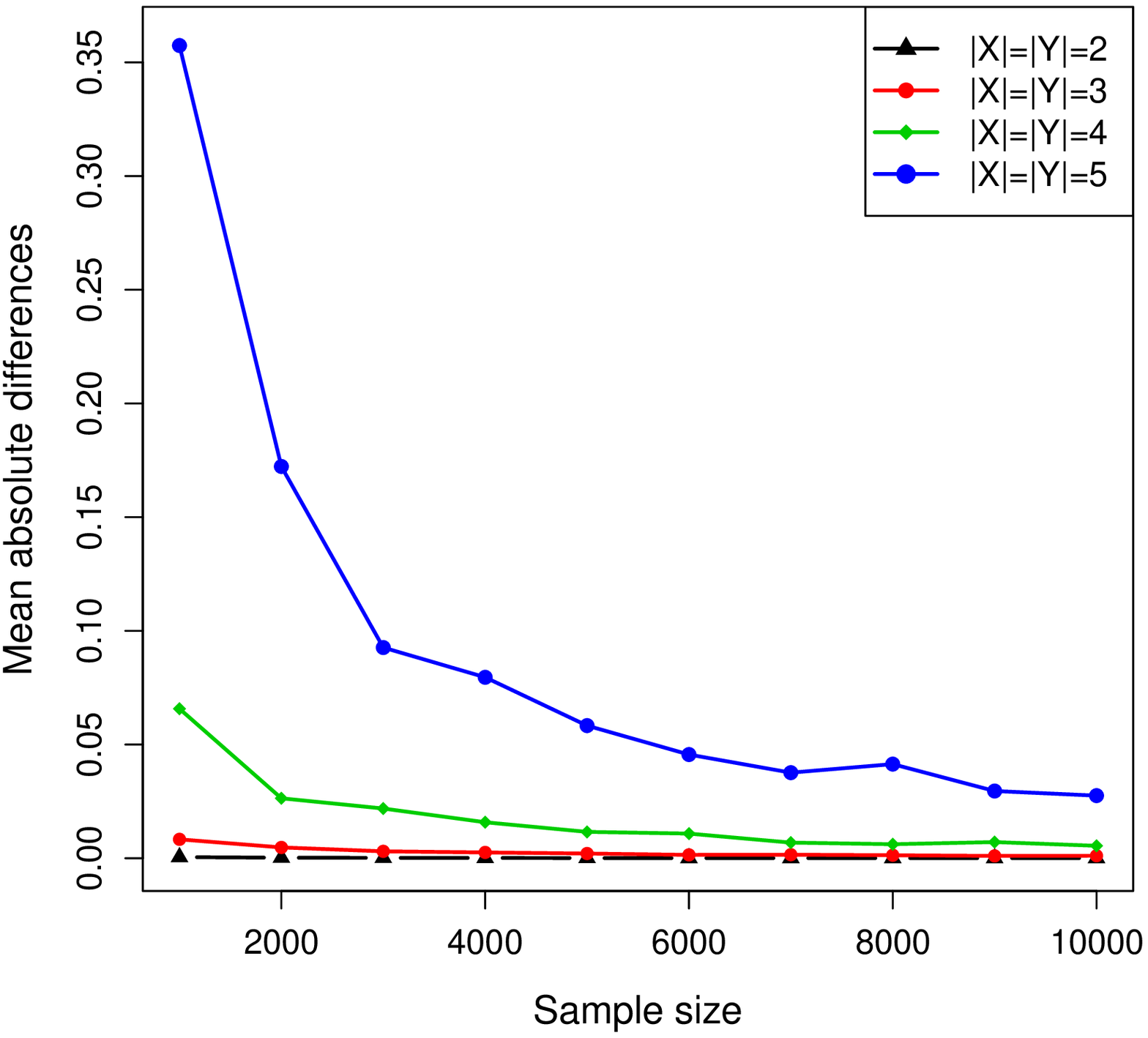} &
\includegraphics[scale = 0.32, trim = 35 0 0 0]{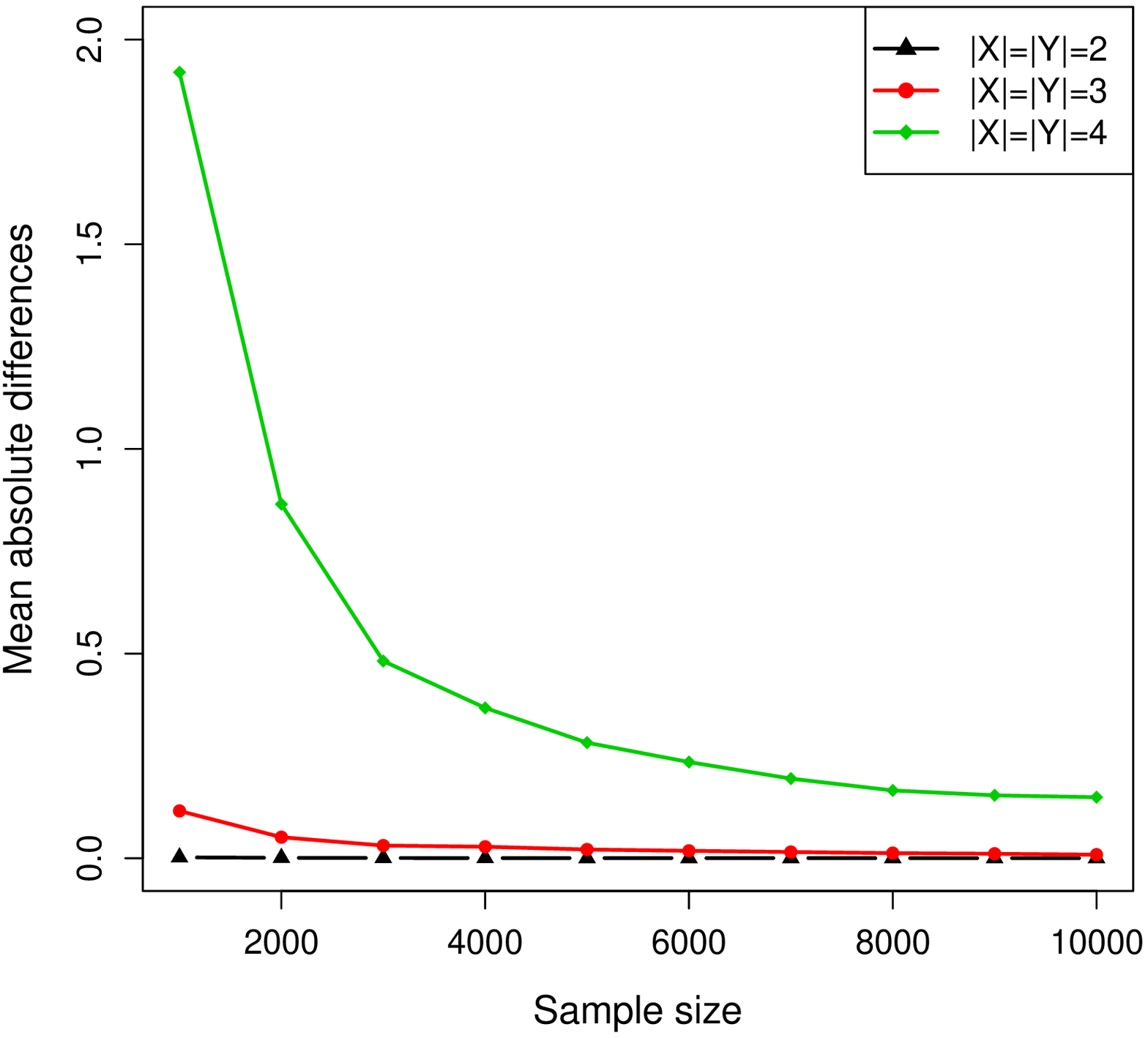} & 
\includegraphics[scale = 0.32, trim = 45 0 0 0]{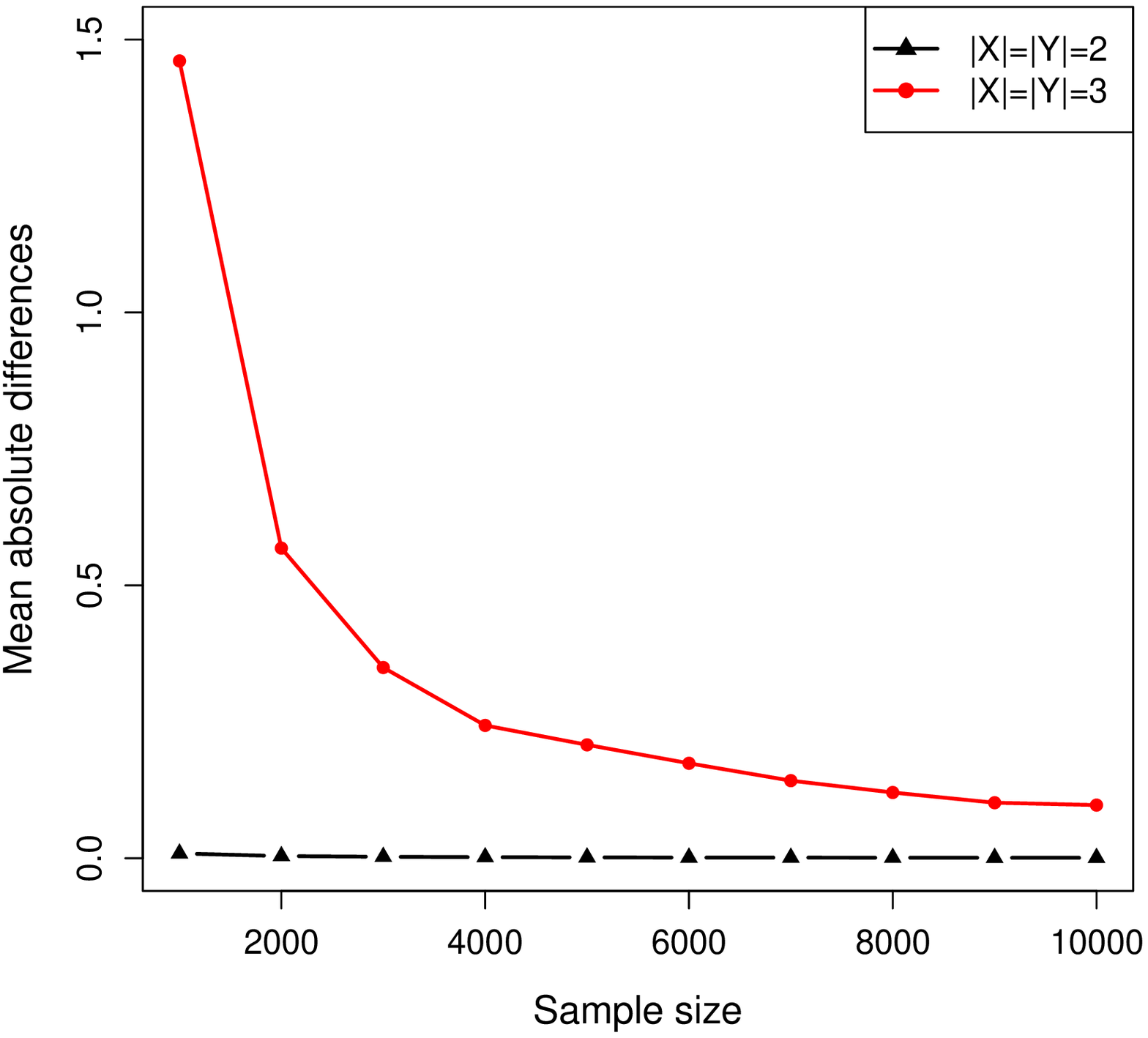} \\
(d)    &    (e)   &    (f)  \\
\textbf{No conditioning variables}  &  \textbf{1 conditioning variable}  & \textbf{2 conditioning variables}  \\
       &          &        
\end{tabular}
\caption{Average difference between the $G^2$ and the $X^2$ ($G^2 - X^2$) test statistics. The range of sample sizes (in the first row goes up to $1,000$, each time increasing by a step equal to $20$. The range of sample sizes in the second row varies from $1,000$ up to $10,000$ increasing by a step equal to $1,000$. The first column refers to unconditional independence, while the second and third columns refer to conditional independence with 1 and 2 conditioning variables respectively. The cardinalities of the variables are showed with different colours. In the first row, the $X^2$ could not be computed for all cases with 1 and 2 conditioning variables. In order to be consistent, in the second row, the the $X^2$ was not be computed for all cases either. \label{accuracy} }
\end{figure}

Figure \ref{accuracy} illustrates the average difference between the two test statistics $G^2 - X^2$ for a range of sample sizes, in the case of unconditional independence. Figure \ref{accuracy} contains the differences for small sample sizes, up to $1,000$, and for larger sample sizes, from $1,000$ up to $10,000$. The $G^2$ test statistic is on average greater than the $X^2$ test statistic and the differences are more pronounced as the cardinalities of the variables increase. This figure clearly shows that the $X^2$ approximation to the $G^2$ test statistic requires larger sample sizes with increasing cardinalities. In all cases though, the differences decay towards zero as the sample size is at the order of hundreds. Yet, the differences between the two test statistics are large and this explains the inflated type I error of the $G^2$ test, observed later. 

A very interesting finding is that as the number of conditioning variables increases, the $X^2$ could not be computed because of zero expected values. Our data generation mechanism is based on the binomial distribution and hence the produced tables will contain zero rows and or zero columns. Applying the $X^2$ test on such tables results in division of a finite number with 0 yielding infinite numbers and hence the $X^2$ is not applicable. In contrast, the $G^2$ test is applicable since $0log{0}=0$ which is a finite number. 

\subsection{Type I error}
At first we will consider the type I error of the three hypothesis testing procedures in four case scenarios. When both $X$ and $Y$ take 2 values, 3 values, 4 values and 5 values. These correspond to $2 \times 2$, $3 \times 3$, $4 \times 4$ and $5 \times 5$ contingency tables respectively. At first, we considered the type I error for the unconditional independence case scenario when the data were generated from a discrete uniform distribution $U(0, i)$, where $i=1,2,3,4$. This is a small case scenario and we did not examine it in more detail. We then generated random values from $Bin(i, 0.5)$, where $i=1,2,3,4$ and most of the times yielded contingency tables with zero frequencies. The binomial distribution was the main distribution used thought the simulation studies. In all cases, we generated an $n \times 100$ matrix, where $n$ denotes the sample size and performed all $100*(100-1)/2=4950$ tests with either testing procedure.  

All results presented in Figures and Tables refer to the cases of the sample sizes being at most $1,000$. For larger sample sizes we only tested the $X^2$ and the $G^2$ tests and they exhibited nearly the same performance and were always size correct and hence not visualized. 

\begin{figure}[ht]
\centering
\begin{tabular}{cc}
\includegraphics[scale = 0.33, trim = 0 0 0 0]{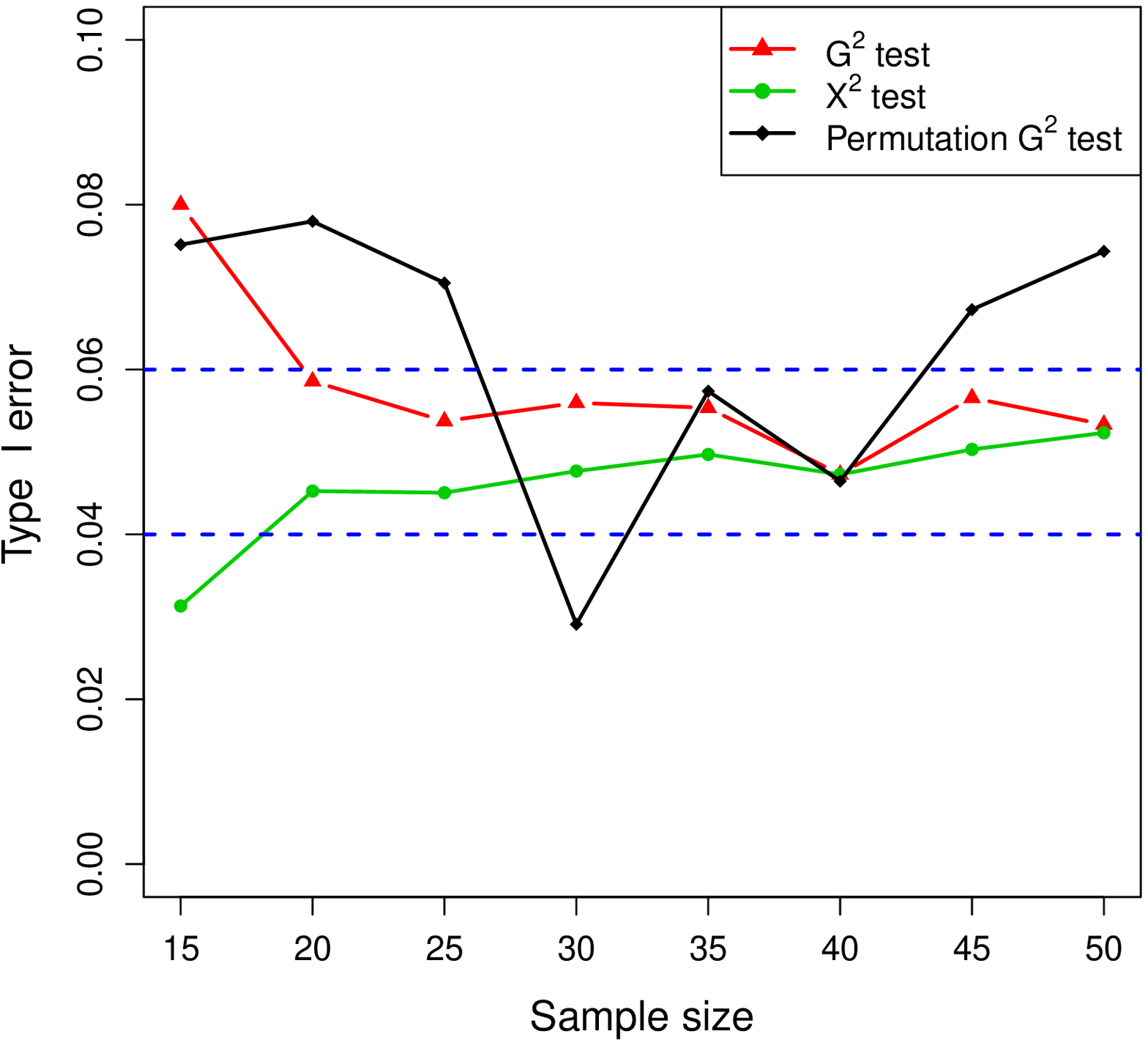}  &
\includegraphics[scale = 0.33, trim = 0 0 0 0]{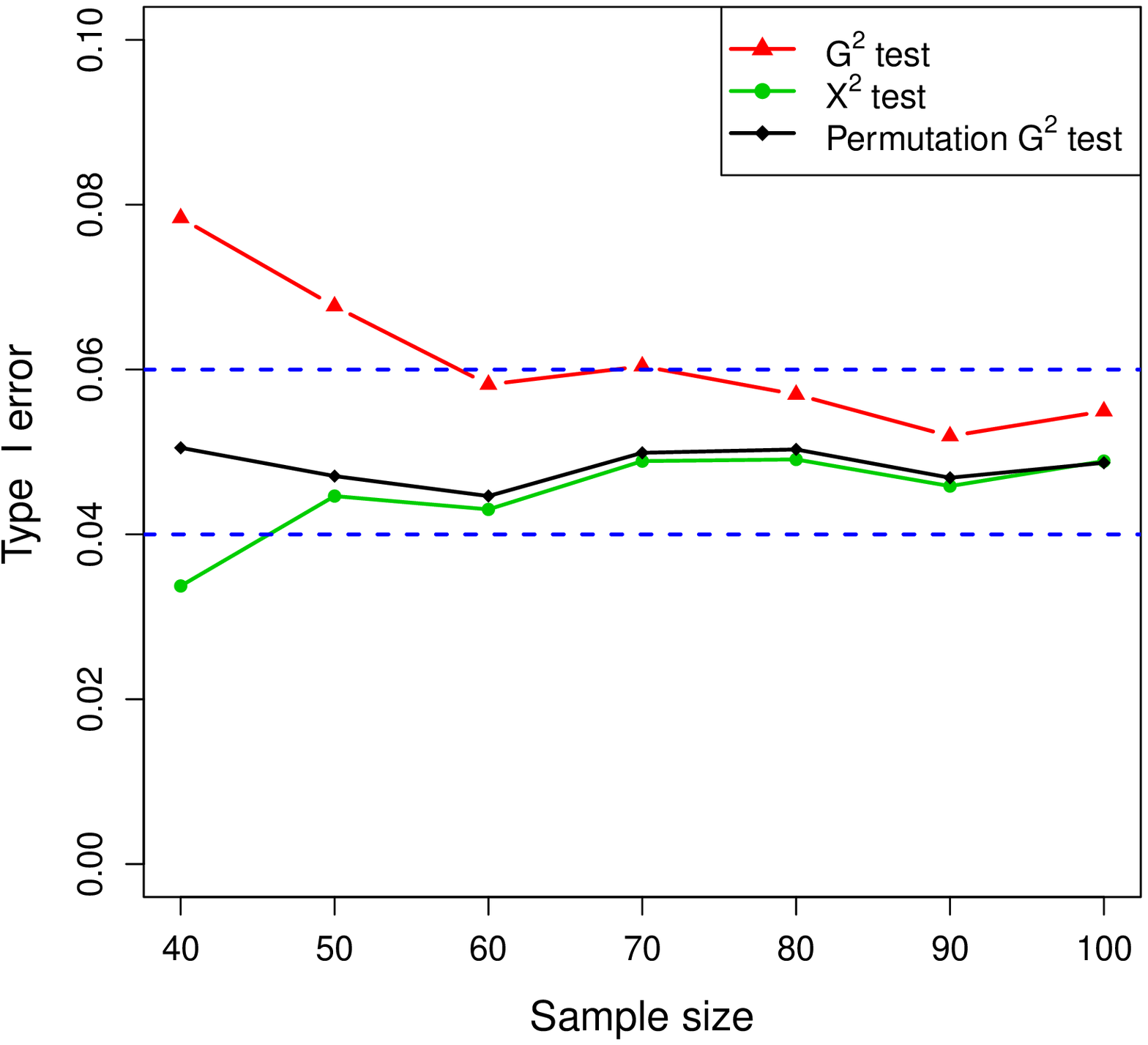}  \\
\textbf{(a) $|X| = |Y| = 2$}   &   \textbf{(b) $|X| = |Y| = 3$}  \\
\includegraphics[scale = 0.33, trim = 0 0 0 0]{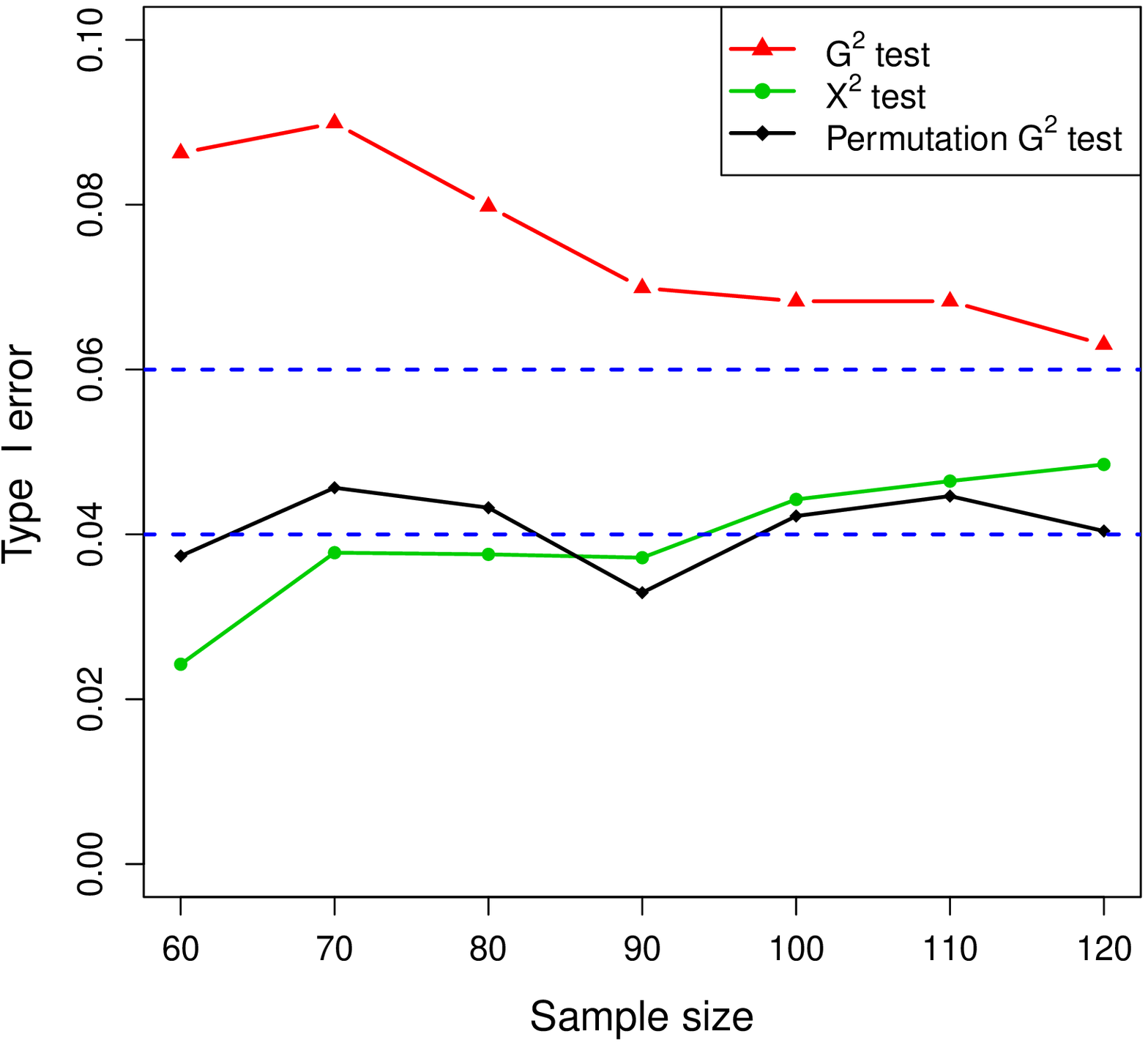}  & 
\includegraphics[scale = 0.33, trim = 0 0 0 0]{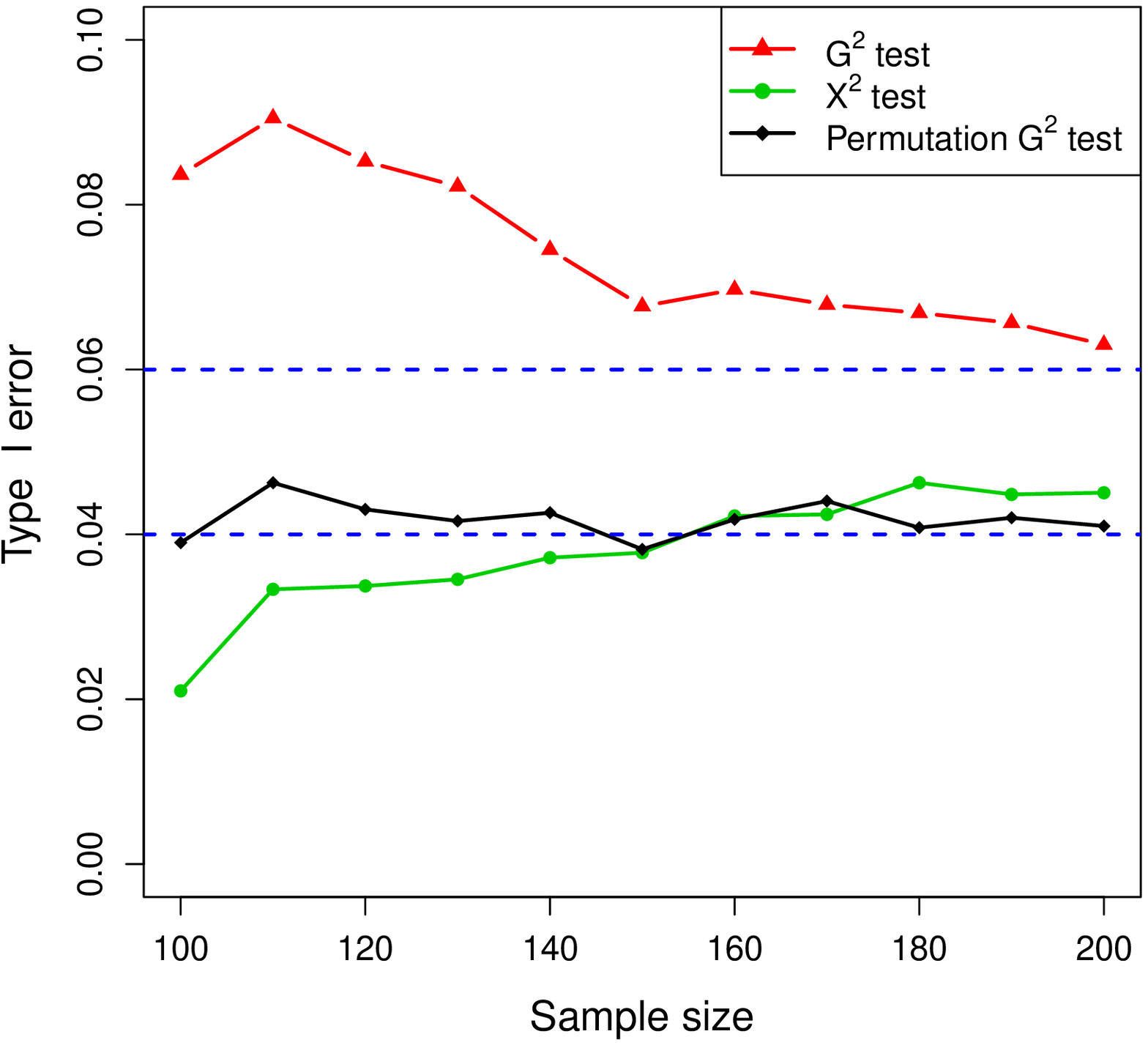}  \\
\textbf{(c) $|X| = |Y| = 4$}   &   \textbf{(d) $|X| = |Y| = 5$}  \\
\end{tabular}
\caption{Estimated type I error of the $G^2$, permutation $G^2$ and $X^2$ tests of independence as a function of the sample size with no conditioning variable. The data were generated from a discrete uniform distribution and the sample sizes were at most $1,000$. \label{alpha_unif} }
\end{figure}

\begin{table}[!ht]
\caption{Proportion of times a testing procedure attained the type I error in the unconditional independence when the data were generated from a discrete uniform distribution and the sample sizes were at most 1,000. The highest proportions are bolded. \label{tab_unif_alpha}}
\begin{center}
\begin{tabular}{c|ccc} \hline \hline 
Cardinalities  & $X^2$ test  &  $G^2$ test  & Permutation $G^2$ test \\  \hline \hline
$|X| = |Y| = 2$   &  7/8    &  7/8    &  2/8    \\
$|X| = |Y| = 3$   &  6/7    &  5/7    &  7/7    \\
$|X| = |Y| = 4$   &  3/7    &  0/7    &  5/7    \\
$|X| = |Y| = 5$   &  4/11   &  0/11   &  9/11   \\  \hline 
Totals            &  20/33  &  12/33  &  \textbf{23/33}  \\  \hline  \hline
\end{tabular}
\end{center}
\end{table}

\begin{figure}[ht]
\centering
\begin{tabular}{cc}
\includegraphics[scale = 0.33, trim = 0 0 0 0]{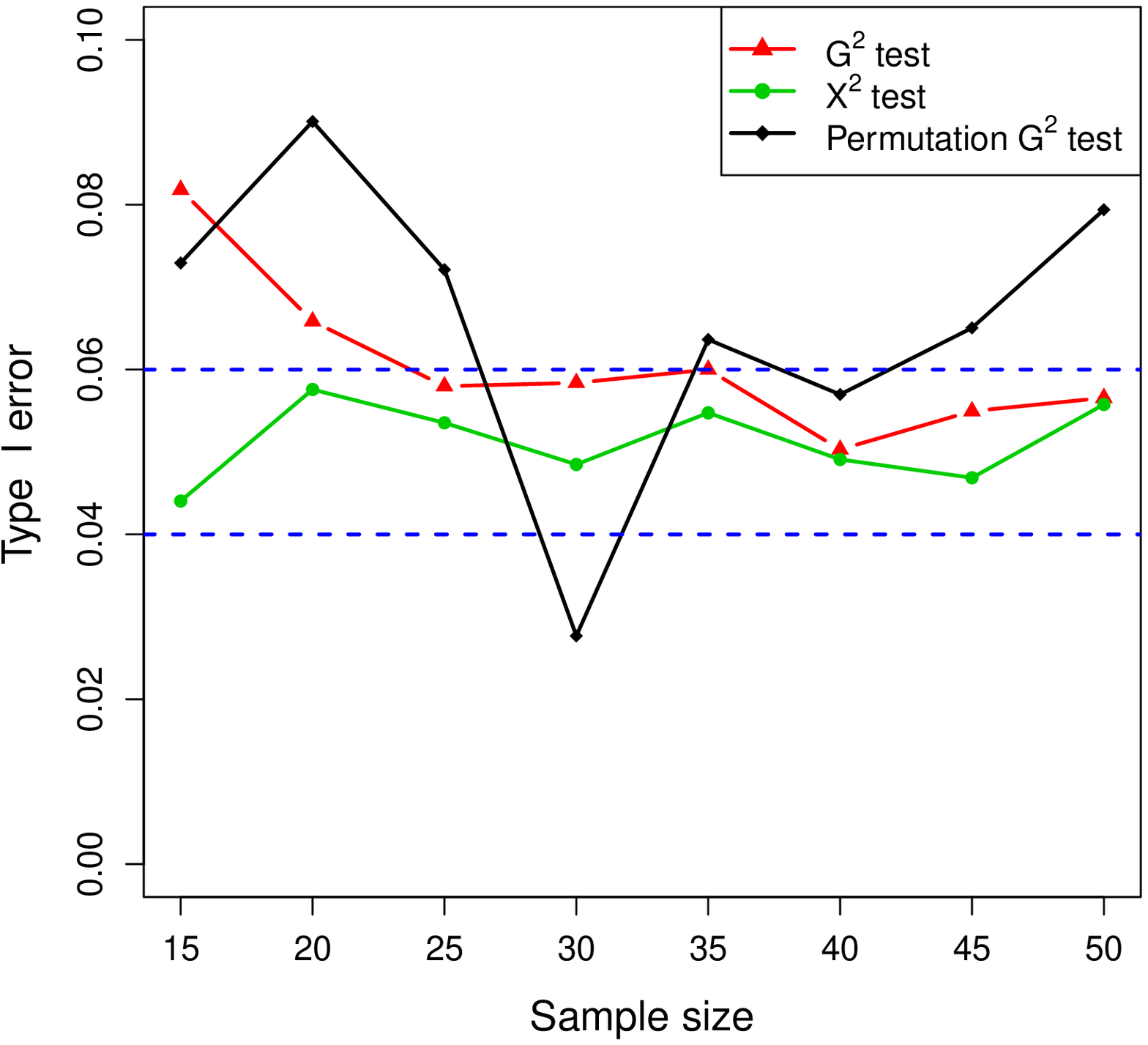}  &
\includegraphics[scale = 0.33, trim = 0 0 0 0]{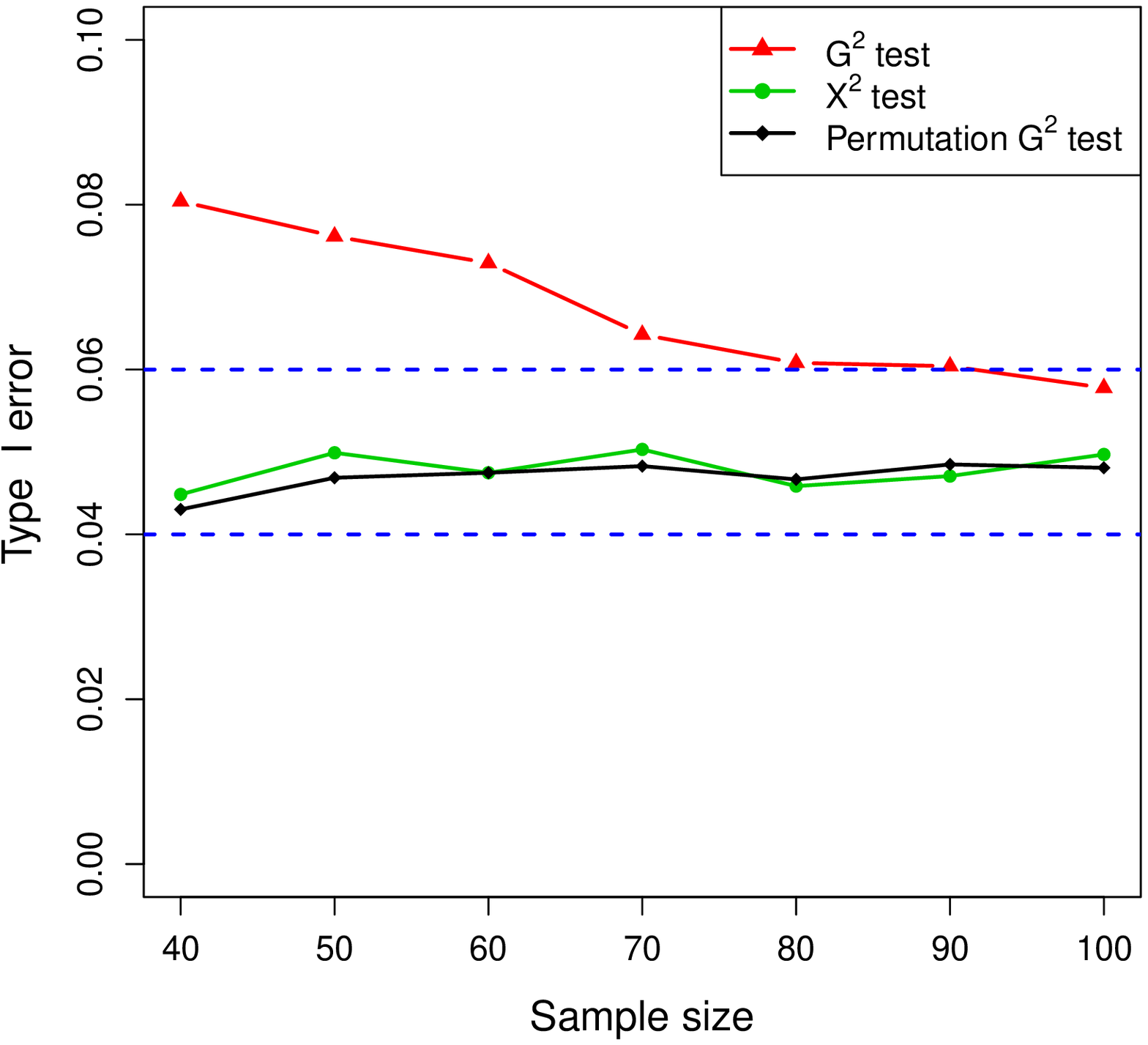}  \\
\textbf{(a) $|X| = |Y| = 2$}   &   \textbf{(b) $|X| = |Y| = 3$}  \\
\includegraphics[scale = 0.33, trim = 0 0 0 0]{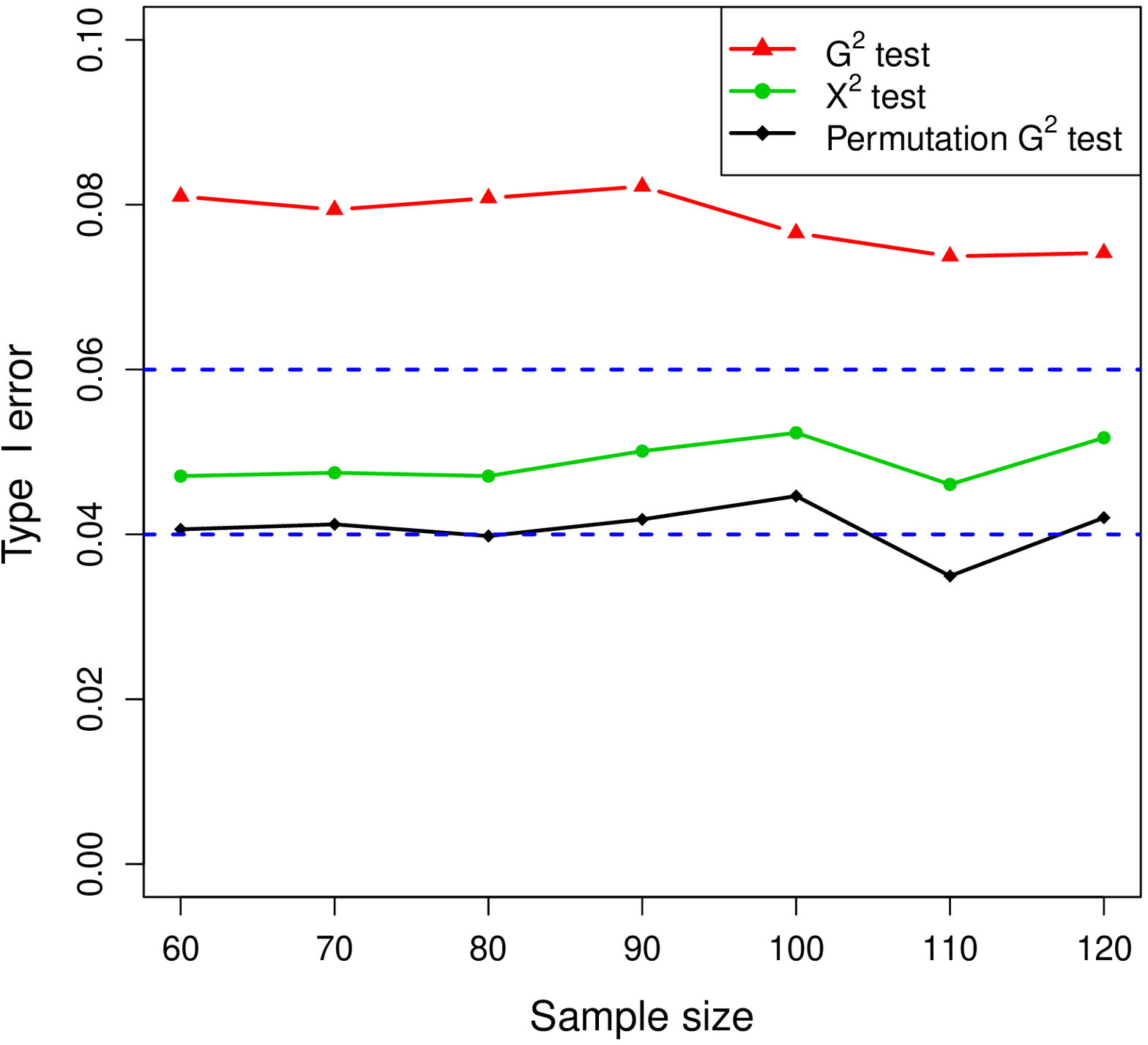}  & 
\includegraphics[scale = 0.33, trim = 0 0 0 0]{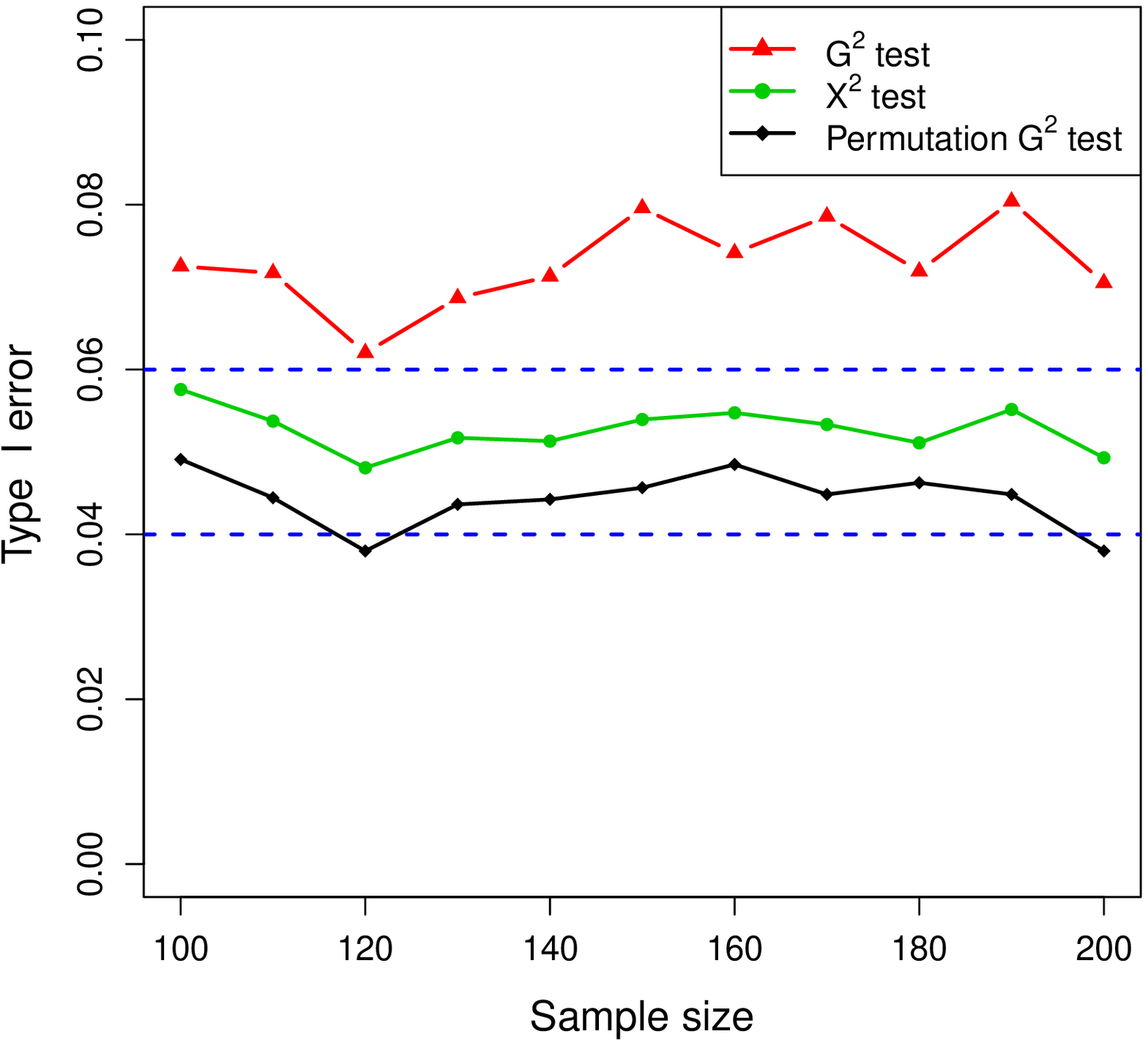}  \\
\textbf{(c) $|X| = |Y| = 4$}   &   \textbf{(d) $|X| = |Y| = 5$}  \\
\end{tabular}
\caption{Estimated type I error of the $G^2$, permutation $G^2$ and $X^2$ tests of independence as a function of the sample size with no conditioning variable. The data were generated from a binomial distribution and the sample sizes were at most $1,000$. \label{alpha_0} }
\end{figure}

\begin{figure}[ht]
\centering
\begin{tabular}{cc}
\includegraphics[scale = 0.33, trim = 0 0 0 0]{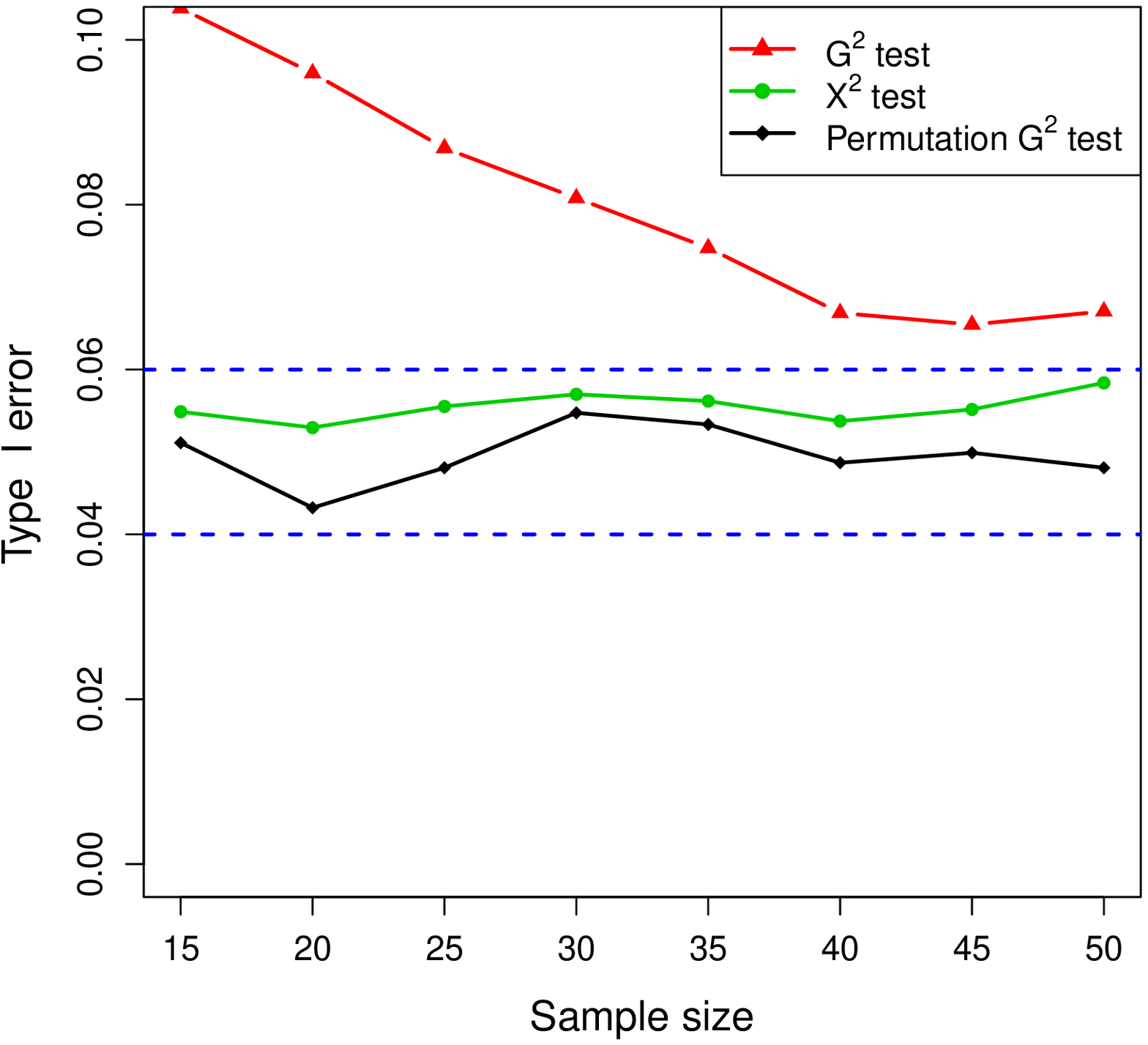}  &
\includegraphics[scale = 0.33, trim = 0 0 0 0]{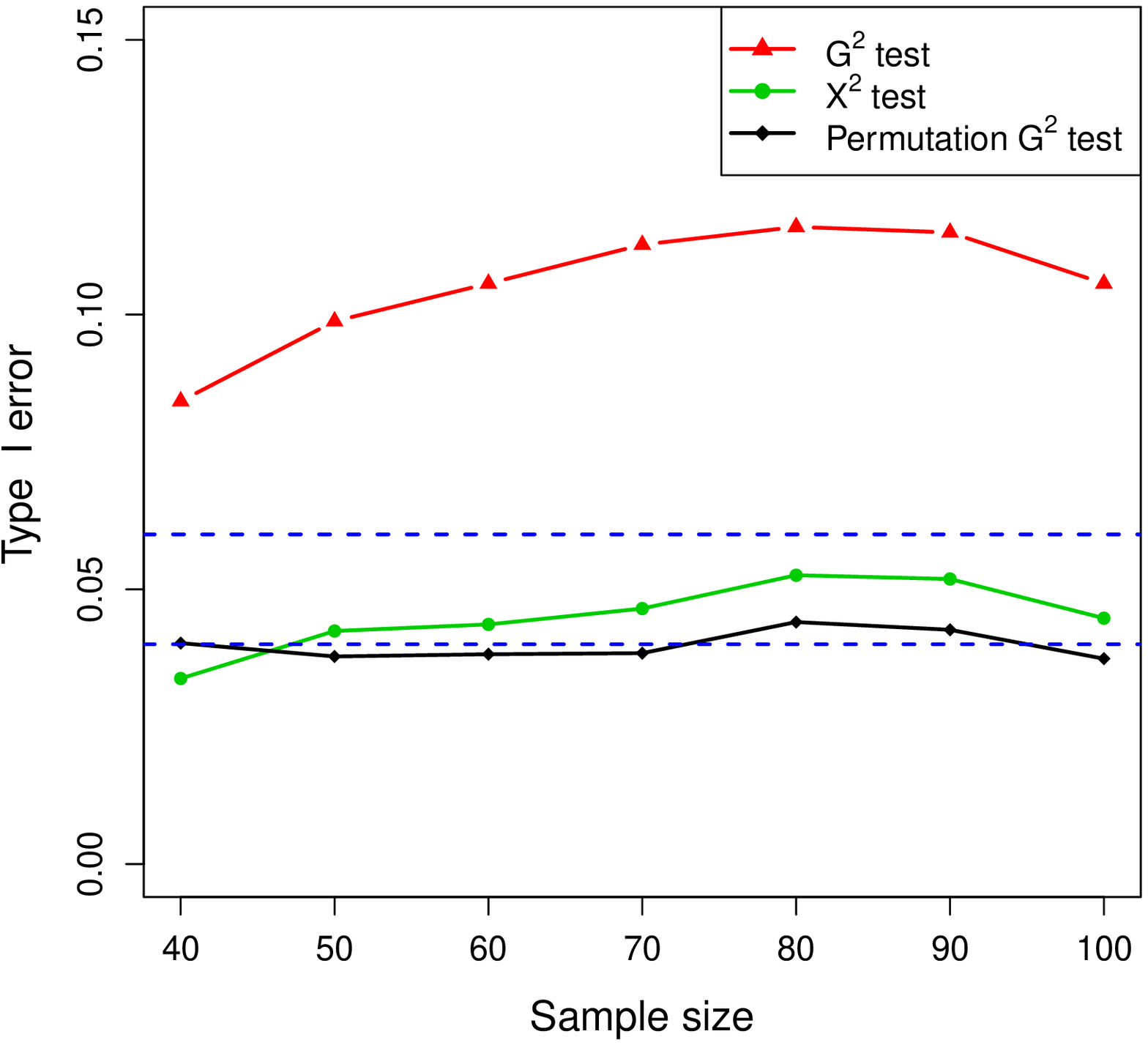}  \\
\textbf{(a) $|X| = |Y| = 2$}   &   \textbf{(b) $|X| = |Y| = 3$}  \\
\includegraphics[scale = 0.33, trim = 0 0 0 0]{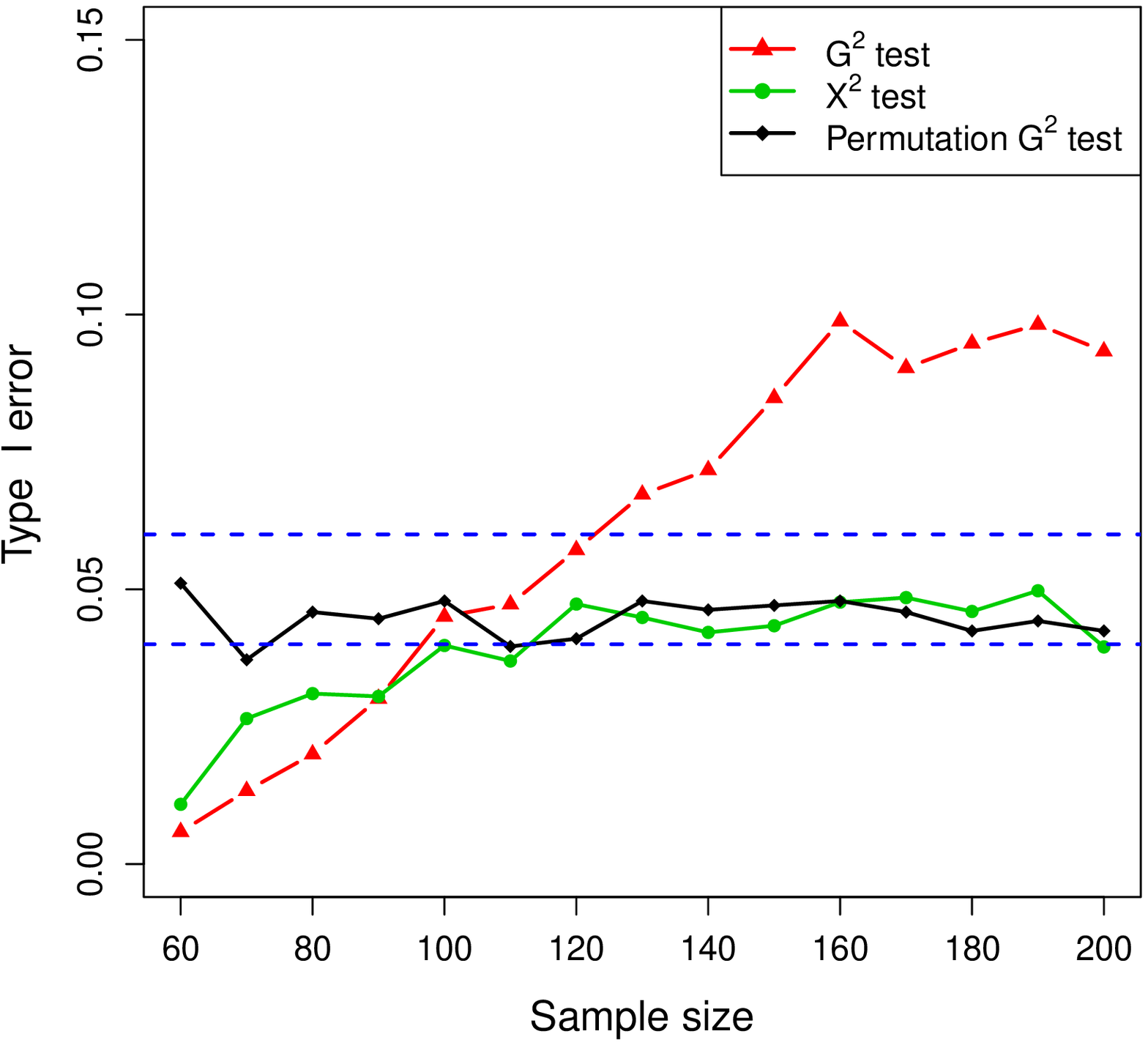}  & 
\includegraphics[scale = 0.33, trim = 0 0 0 0]{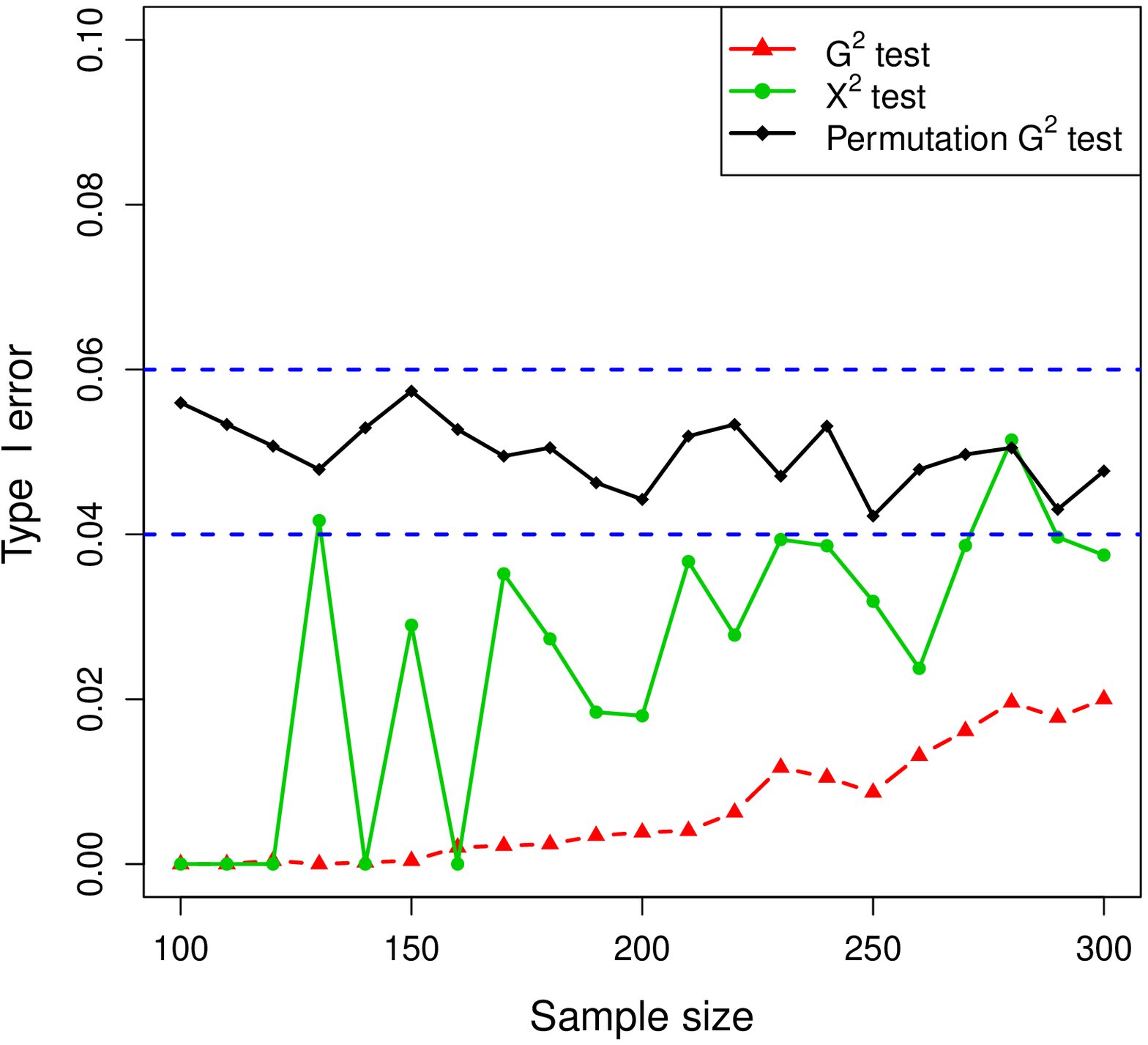}  \\
\textbf{(c) $|X| = |Y| = 4$}   &   \textbf{(d) $|X| = |Y| = 5$}  \\
\end{tabular}
\caption{Estimated type I error of the $G^2$, permutation $G^2$ and $X^2$ tests of independence for different cardinalities as a function of the sample size with 1 conditioning variable. The data were generated from a binomial distribution and the sample sizes were at most $1,000$. \label{alpha_1} }
\end{figure}

\begin{figure}[ht]
\centering
\begin{tabular}{cc}
\includegraphics[scale = 0.33, trim = 0 0 0 0]{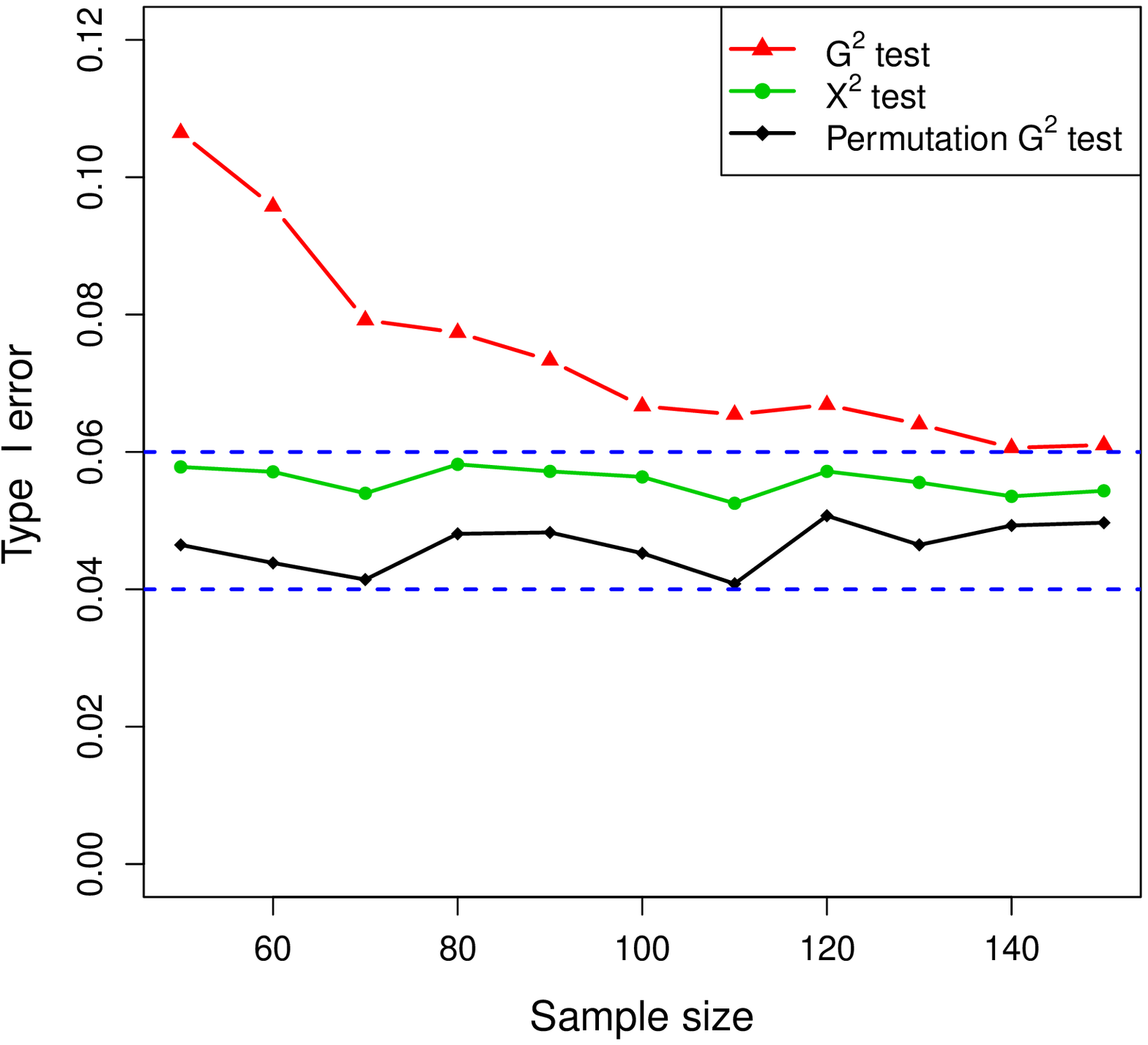}  &
\includegraphics[scale = 0.33, trim = 0 0 0 0]{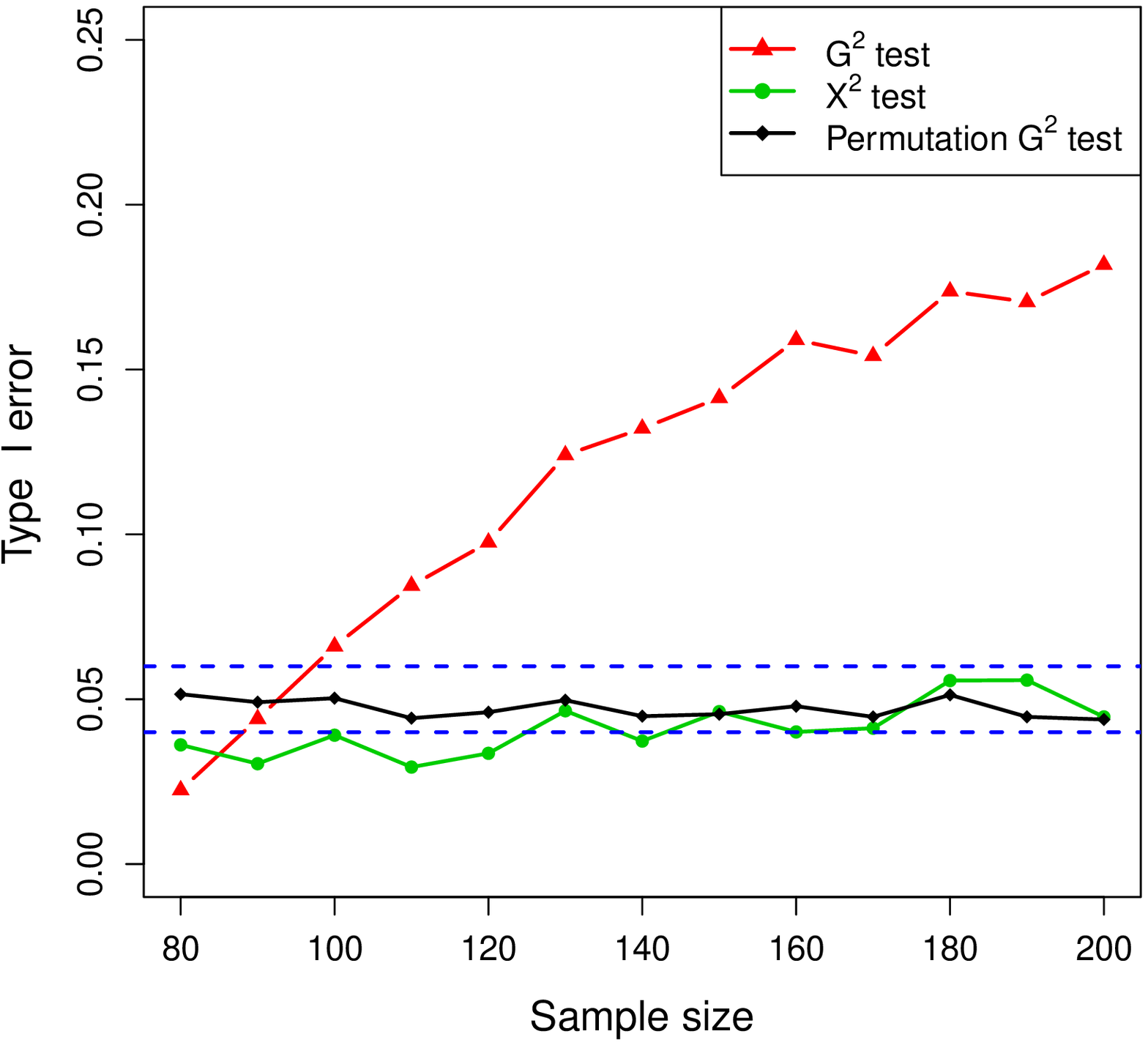}  \\
\textbf{(a) $|X| = |Y| = 2$}   &   \textbf{(b) $|X| = |Y| = 3$}  \\
\includegraphics[scale = 0.33, trim = 0 0 0 0]{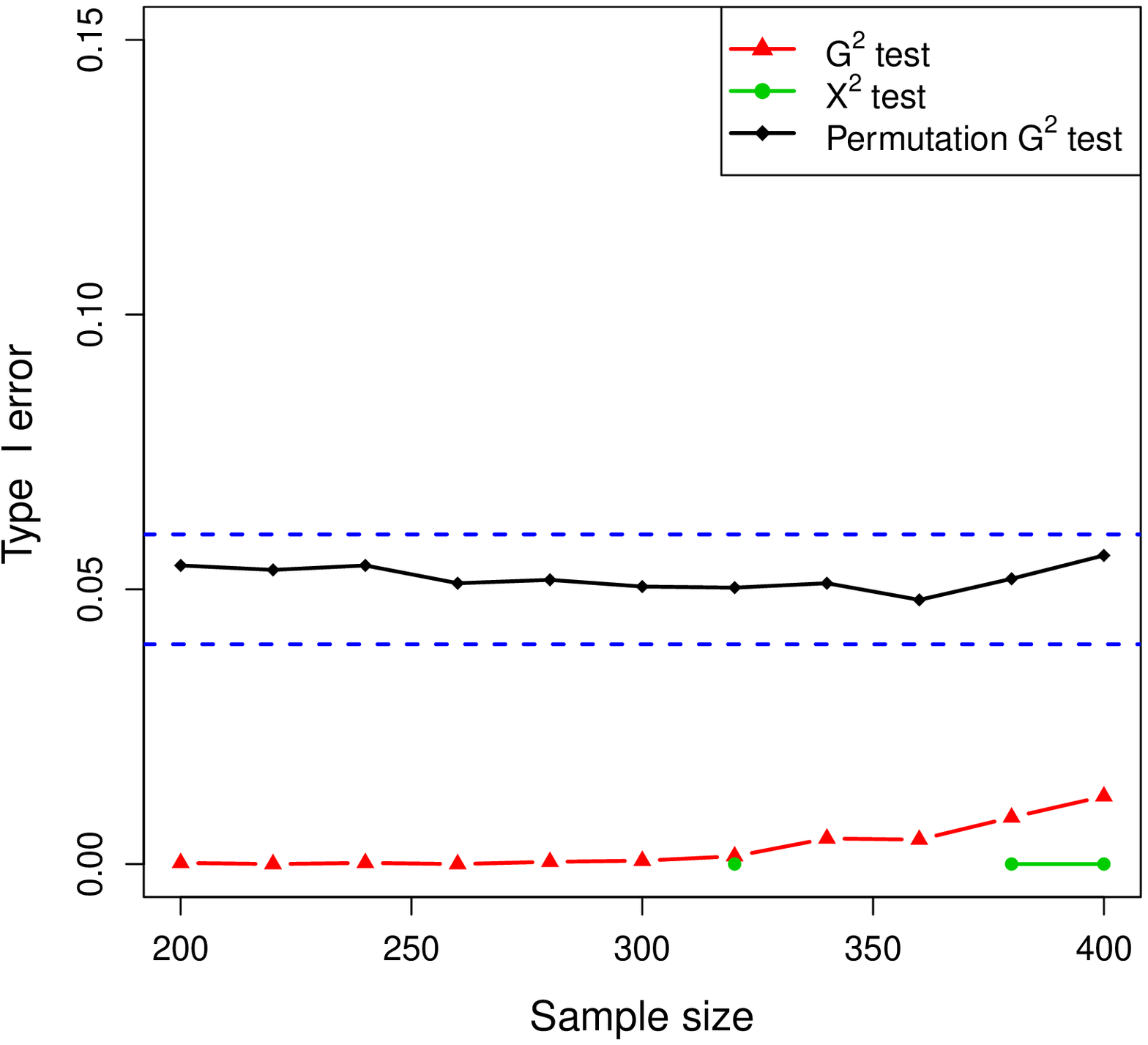}  & 
\includegraphics[scale = 0.33, trim = 0 0 0 0]{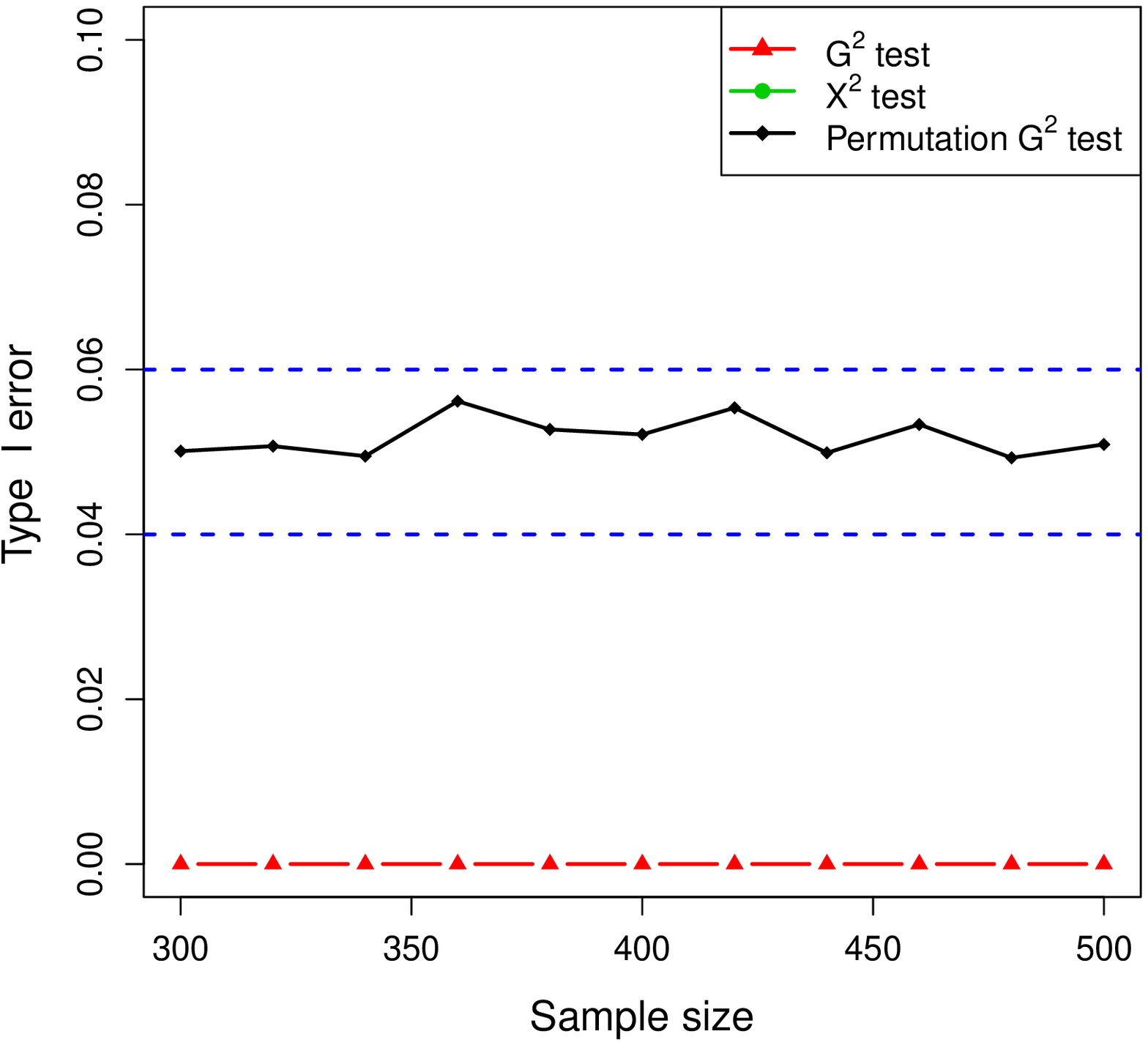}  \\
\textbf{(c) $|X| = |Y| = 4$}   &   \textbf{(d) $|X| = |Y| = 5$}  \\
\end{tabular}
\caption{Estimated type I error of the $G^2$, permutation $G^2$ and $X^2$ tests of independence for different cardinalities as a function of the sample size with 2 conditioning variables. The data were generated from a binomial distribution and the sample sizes were at most $1,000$. \label{alpha_2} }
\end{figure}

Figure \ref{alpha_0} illustrates the estimated type I error of all testing procedures when testing the unconditional independence of two variables, for various sample sizes and cardinalities of the variables. The most accurate is the $X^2$, followed by the permutation $G^2$, whereas in all cases, except for Figure \ref{alpha_0}(a), the $G^2$ test shows an inflated type I error.  

By examining Figure \ref{alpha_1} we can see that the conclusions change. The $G^2$ test exhibits the worst performance in terms of type I error, as it is either too conservative, i.e. it rejects less frequently than it should, or it is very risky as it tends to reject more frequently. The $X^2$ test seems to perform adequately, but when the cardinalities of the variables are equal to 5, it becomes too conservative. The permutation $G^2$ test on the other hand is the only that holds the type I error within the acceptable limits.  

Finally, Figure \ref{alpha_2} yields different conclusions again. The permutation $G^2$ test still remains size correct, but the $X^2$ test performs adequately only in two cases, with $|X|=|Y|=2$ and $|X|=|Y|=3$. In the other two cases, it could not be computed, due to the conditional contingency tables formed, that contain rows and or columns with zero frequencies ,and hence the $X^2$ test statistic could not be computed. the $G^2$ test on the other hand almost never rejects the conditional independence.  

Among the three testing procedures, the $G^2$ manifests the worst performance in terms of attaining the type I error as it was size correct in only 11 out 130 times (8.46\%), as depicted in Table \ref{tab_alpha}. The $X^2$ test performed better as it was size correct in 76 out 130 times (58.46\%), whereas the permutation $G^2$ test was size correct in 120 out of 130 times (92.31\%). If we examine Table \ref{tab_alpha} more closely, we can draw more specific and more targeted conclusions. Specifically, when testing the (unconditional) independence between two variables, the following can be said. The $X^2$ test exhibited the best performance, followed by the permutation $G^2$ test. The $G^2$ test had the worst performance, as it was size correct in almost 1 out of 4 times.  

At this point let us remind the reader about the necessary condition the expected values should satisfy. The number of cells whose expected values is less than 5 should be no more than 20\%-25\% and no cell must contain 0 frequencies. According to \cite{mchugh2013} \cite{mchugh2013}, no cell should contain expected values that are less than 3. This does not hold true for the $X^2$ test and this is clearly observed in Figure \ref{alpha_0}, where we can see that even for small sample sizes, the $X^2$ retains the nominal significance level (5\%). When we move downwards Table \ref{tab_alpha} we draw another interesting conclusion. The permutation $G^2$ test, unlike the $X^2$ test is size correct in the majority of the cases.

\begin{table}[!ht]
\caption{Proportion of times a testing procedure attained the type I error for the different cases when the data were generated from a binomial distribution and the sample sizes were at most 1,000. The highest proportions are bolded. \label{tab_alpha}}
\begin{center}
\begin{tabular}{c|c|ccc} \hline \hline 
\# of conditional variables & Cardinalities  & $X^2$ test  &  $G^2$ test  & Permutation $G^2$ test \\  \hline \hline
\multirow{4}{*}{$Z=0$} & $|X| = |Y| = 2$   &  8/8    &  6/8   &  6/8   \\
                       & $|X| = |Y| = 3$   &  7/7    &  1/7   &  7/7  \\
                       & $|X| = |Y| = 4$   &  7/7    &  0/7   &  6/7   \\
                       & $|X| = |Y| = 5$   &  11/11  &  0/11  &  9/11  \\ \hline 
               Totals  &                   &  \textbf{33/33}  &  7/33  &  28/33  \\ \hline \hline
\multirow{4}{*}{$Z=1$} & $|X| = |Y| = 2$   &  8/8    &  0/8   &  8/8   \\
                       & $|X| = |Y| = 3$   &  6/7    &  0/7   &  3/7  \\
                       & $|X| = |Y| = 4$   &  9/15   &  3/15  &  14/15   \\
                       & $|X| = |Y| = 5$   &  2/21   &  0/21  &  21/21  \\ \hline 
               Totals  &                   &  25/51  &  3/51  &  \textbf{46/51}  \\ \hline \hline
\multirow{4}{*}{$Z=2$} & $|X| = |Y| = 2$   &  11/11  &  0/11  &  11/11   \\
                       & $|X| = |Y| = 3$   &  7/13   &  1/13  &  13/13  \\
                       & $|X| = |Y| = 4$   &  0/3    &  0/11  &  11/11   \\
                       & $|X| = |Y| = 5$   &  0/11   &  0/11  &  11/11  \\ \hline 
               Totals  &                   &  18/46  &  1/46  &  \textbf{46/46}  \\ \hline \hline 
\end{tabular}
\end{center}
\end{table}

\subsection{Power}
Power comparisons of testing procedures are meaningful only for tests that are size correct. For this reason, we should compare only the $X^2$ with the permutation $G^2$ in the case of unconditional independence. We have compared all three tests though in the unconditional independence case scenario. The specific alternatives we used are described as follows.  
\begin{enumerate}
\item For a range of values of $b$, from -3 up to 3, increasing with a step-size equal to 1 do the following steps.
\item Generate $n$ values $x_i$ from a $Bin(|X|, 0.5)$, where $i=1,\ldots,n$ and $|X|$ is the cardinality of the variable $X$.
\item Compute $p_i=\frac{e^{-sign(b) + b x_i}}{1 + e^{-sign(b) + b x_i}}$, where $sign(.)$ is the sign function that returns the sign of a number.
\item Generate $n$ values $y_i$ from $Bin(|Y|, p_i)$, where $|Y|$ is the cardinality of the variable $Y$ and $|Y| = |X|$.
\item Repeat steps 2-4 $1000$ times. 
\item Compute the power as the proportion of times the $H_0$ is rejected.
\end{enumerate}

\begin{figure}[ht]
\centering
\includegraphics[scale = 1.7, trim = 20 0 0 0]{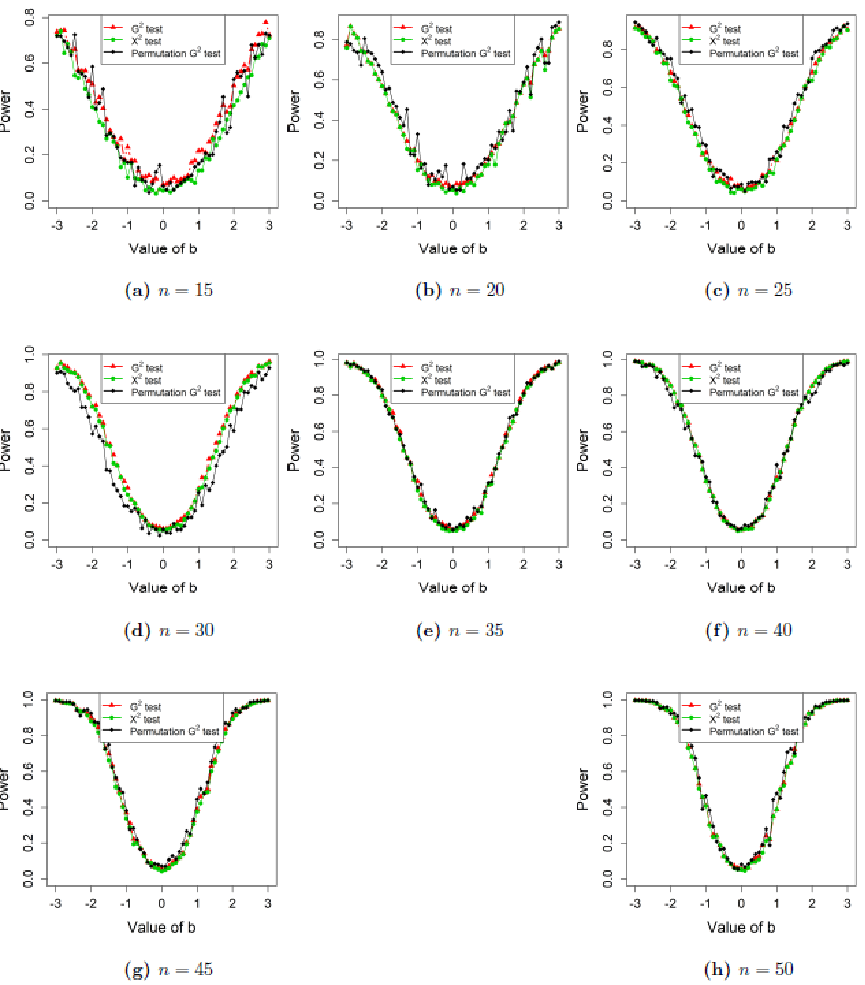}  
\caption{Estimated power of the $G^2$, permutation $G^2$ and $X^2$ tests of independence as a function of the sample size  when $|X|=|Y|=2$ and with no conditioning variable. The data were generated from a binomial distribution and the sample sizes were at most $1,000$. \label{pow2} }
\end{figure}

\begin{figure}[ht]
\centering
\includegraphics[scale = 1.7, trim = 20 0 0 0]{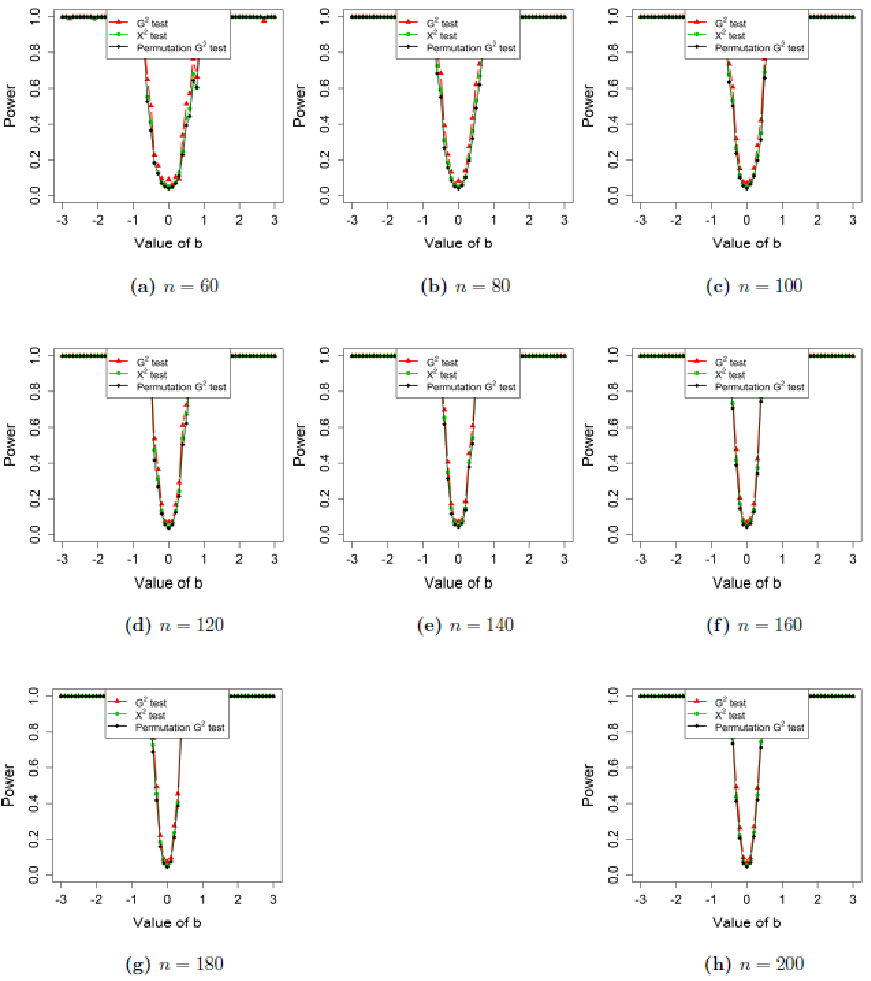}  
\caption{Estimated power of the $G^2$, permutation $G^2$ and $X^2$ tests of independence as a function of the sample size  when $|X|=|Y|=4$ and with no conditioning variable. The data were generated from a binomial distribution and the sample sizes were at most $1,000$. \label{pow4} }
\end{figure}

Figures \ref{pow2} and \ref{pow4} visualize the estimated powers for a range of different alternatives when the cardinalities of the variables are equal to 2 and 4. The powers are similar for the other cases and hence omitted for brevity. We observe that the estimated power levels are similar for all three tests. We emphasize that the power of the $G^2$ presented in Figure \ref{pow4} is not comparable to the powers of the $X^2$ test and the parametric $G^2$ test because it was not size correct (see Figures \ref{alpha_0} - \ref{alpha_2} and Table \ref{accuracy}).  

\section{Conclusions}
We performed Monte Carlo simulation studies aiming to provide evidence as to the suitability of two popular tests for independence of categorical variables. The simulations proved useful and we drawn very interesting conclusions. We highlight that the following conclusions apply to sample sizes of at most equal to $1,000$.
 
When testing the (unconditional) independence of two variables, the $G^2$ test never attained the type I error when the variables contain 3 or more values (or levels). When the variables take 3 values, the $G^2$ test was size correct only for large sample sizes. On the contrary, the $X^2$ and the permutation $G^2$ test were size correct in almost all examined cases, regardless of the sample size and of the number of values the categorical variables could take. This is a strong evidence that the rule of thumb regarding the unsuitability of the $X^2$ test with zero value frequencies and/or expected values being less than 5 does not hold true and should be carefully re-examined. Based on our simulations, we have some ground to say that the rule's validity seems to be rather small or negligible.  

When testing conditional independence of two variables, with 1 or 2 conditioning variables, the permutation $G^2$ test was the only one among the three competitors that performed satisfactorily. It was the most accurate, in terms of attaining the type I error, in more than 90\% of the examined cases, followed by the $X^2$ which was size correct in only 58\% of the cases. The $G^2$ test was size correct in less than 10\% of the examined cases. The value of the $G^2$ test statistic was always greater than that of the $X^2$ test, but this difference was diminishing with increasing sample sizes. On the other hand, the $X^2$ could not always be computed due to infinite divisions.  

On the other hand, when the sample sizes are at the orders of thousands and higher, use of the $X^2$ and of the $G^2$ is strongly suggested as they both perform equally well. They are always size correct and poses similar levels of power. Both of them are also computationally extremely efficient, hence we suggest application of either of them with large sample sizes. Computational efficiency and statistical accuracy are two essential components of a test in the current era of massive and big data.


\end{document}